\documentclass[aps,prx,amsmath,amssymb,groupedaddress,superscriptaddress,floatfix,twocolumn,showkeys,amsart]{revtex4-1}
\usepackage{graphicx} 
\usepackage{tabularray}
\usepackage[caption=false]{subfig}
\usepackage{float}
\usepackage{xcolor}
\usepackage{diagbox}

\usepackage{comment}
\usepackage{siunitx}
\usepackage{xr}

\renewcommand{\S}[0]{^1\rm{S}_0}
\renewcommand{\P}[0]{^3\rm{P}_0}
\newcommand{\Pone}[0]{^3\rm{P}_1}
\newcommand{\SP}[0]{\S\leftrightarrow\P}
\newcommand{\Al}[0]{^{27}\rm{Al}^+}
\newcommand{\ket}[1]{|#1\rangle}
\newcommand{\sss}[0]{\scriptscriptstyle}

\begin{document}
\title{Extending the fundamental limit of atomic clock stability}

\author{R.~Shaniv}\affiliation{Time and Frequency Division, National Institute of Standards and Technology, Boulder, CO 80305, USA }
\author{A.~Agrawal}
\affiliation{Department of Physics, University of Oxford, Clarendon Laboratory, Parks Road, Oxford OX1 3PU, United Kingdom}
\author{D.~B.~Hume}\affiliation{Time and Frequency Division, National Institute of Standards and Technology, Boulder, CO 80305, USA }\affiliation{Department of Physics, University of Colorado, Boulder, Colorado, 80309, USA}

\begin{abstract}
Optical atomic clocks have been rapidly developing in recent decades, resulting in major improvements in both precision and accuracy. As a result, they have become instrumental in multiple areas of applied and fundamental research. Despite all atomic frequency references having more than two energy-levels, the commonly used model for evaluating their ultimate limits assumes a two-level atom. This leads to frequency interrogation protocols and theoretical stability bounds that are suboptimal for a true multi-level atom. The most fundamental stability bound assumes two noise sources - quantum projection noise and spontaneous decay from the excited state. In this work, we analyze a model that includes these noise types and is generalized beyond the two-level assumption, where spontaneous decay can branch to more than a single ground state. This model allows for detection and exclusion of atomic frequency interrogations in which the atom decayed, leading to a frequency stability improvement of up to $\approx 4.5 \text{ dB}$ compared with the two-level model. Furthermore, we identify an even greater stability enhancement of $\approx 5.4 \text{ dB}$ for frequency comparisons between atoms in an odd parity Bell state. These enhancements are particularly relevant for the numerous trapped-ion optical clock species that operate close to lifetime-limited stability. We calculate new stability limits for those cases and provide a detailed experimental protocol for frequency interrogation with an $^{27}\text{Al}^{+}$ optical ion clock.
\end{abstract}

\maketitle

\par Optical atomic clocks are among the most precise and accurate measurement devices ever developed. These clocks operate by probing the optical resonance frequency needed to excite an atomic reference from a low energy state to a state with higher energy \cite{poli2013optical,ludlow2015optical,fortier2026optical}. In addition to accurate time keeping and frequency standards \cite{margolis2010optical,riehle2015towards,ma2025absolute,fortier2026optical}, optical atomic references can be used for technological applications such as geodesy \cite{mehlstaubler2018atomic,grotti2024long}, navigation \cite{burt2021demonstration, roslund2024optical}, quantum networks \cite{nichol2022elementary,komar2014quantum} and fundamental research \cite{chou2010optical,peik2021optical,schkolnik2022optical,derevianko2022fundamental}. As a consequence, improving the stability of optical atomic clocks remains an active area of research that can significantly impact technology and science exploration \cite{king2022optical,aeppli2024clock,akerman2025operating,marshall2025high}.

\par Typically, atomic frequency references are modeled as a two-level open quantum system \cite{riis2004optimum,peik2005laser}. The validity of a model reduced to two atomic levels stems from the use of a narrow linewidth probe laser, which addresses a specific atomic transition while leaving dynamics of non-resonant transitions negligible \cite{tannoudji1991atom}. The openness of the model, at minimum, comes from the interaction of the atom with the environment's electromagnetic vacuum, leading to a finite lifetime of any excited state. As probing the transition frequency requires a superposition state, therefore populating an excited state, a fundamental limit is imposed on the stability of the atom due to the combination of quantum projection noise and the excited state lifetime. The former is a consequence of the discreteness of the atomic energy levels \cite{bergquist1994laser,riis2004optimum,peik2005laser} and necessitates averaging the results of repeated experimental frequency interrogations. The latter affects the stability as follows: when in the excited state, an atom can undergo spontaneous emission, leaving it in a lower energy level. Since light is emitted to the environment, the atom collapses to an energy eigenstate and any coherent superposition information is lost. The stochastic nature of this phenomenon, commonly referred to as ``spontaneous'' decay, and the inability to effectively determine whether the atom decayed or not, degrade the stability of the atom's transition frequency inference. Explicitly, all the results from experimental frequency interrogations must be averaged together, despite some of them contributing projection noise but adding no frequency information.

\par The optimal stability within the aforementioned two-level model, referred to as the lifetime limit, has been considered a fundamental bound for protocols relying on uncorrelated atomic references. The experimental parameters corresponding to this limit for both the initial superposition amplitudes and the interrogation time have been derived \cite{bergquist1994laser,riis2004optimum,peik2005laser} and a lifetime-limited operation has been demonstrated \cite{clements2020lifetime,lindvall202588}. 
\par However, all real atomic references possess more than two levels. Although the interrogation primarily addresses two atomic levels, the atom-environment interaction is broader and can affect the entire atomic level structure. Incorporating additional levels as possible outcomes of an interrogation protocol enables the retrieval of information regarding the atomic evolution that is otherwise lost. Similar concepts are applied in quantum information processing for error correction and detection protocols, where an error can be identified if the resulting quantum state lies outside the designated qubit code subspace \cite{nielsen2010quantum}.  


\par Related ideas of employing multi-level detection schemes for atomic lattice clocks have been proposed and implemented \cite{ma2025enhancing}, enhancing the interrogation time when limited by off-resonant Raman scattering from the lattice lasers. In addition, a protocol for reaching beyond the two-level lifetime limit was recently proposed \cite{kielinski2024ghz}, where a cleverly-chosen readout protocol for Greenberger-Horne-Zeilinger (GHZ) states of two-level atoms was considered, leading to improvement in the stability. In that proposal, the gain was attributed to a quantum enhancement and required entangling operations on multiple atoms for both state preparation and detection.

\par In this work, we generalize the two-level model for a single atomic reference to include spontaneous decay that can leave the atom in multiple levels. We show that spontaneous decay can be detected to the extent allowed by the branching ratio, which can be used to increase the clock stability compared to the two-level lifetime limit. In the case of an atomic excited state with negligible branching ratio to the ground state of the clock transition, we show the same stability gain obtained in \cite{kielinski2024ghz} without the need for multiple atoms or entangled states (analysis regarding stability improvement arising from both the multi-level atom model together with GHZ states can be found in the supplementary material). In addition, we extend our analysis to the interrogation of multi-atom symmetric superposition states, e.g. odd parity Bell states. Such states are insensitive to laser decoherence, which is the limiting decoherence source for most atomic clocks, and are particularly useful for clock comparisons, reaching the lifetime limit \cite{clements2020lifetime,zhiqiang2023176lu+,nichol2022elementary}. Our analysis indicates that the stability enhancement for these symmetric states when comparing the presented multi-level model to the two-level model is even greater than that of the single atom case. Further, we introduce and analyze the notion of mid-interrogation decay detection, which allows to restart the interrogation following spontaneous decay without the need to wait the full interrogation time. This enables the performance of more interrogations within a given time, improving stability even further and establishing a new optical atomic clock lifetime limit. Finally, we provide a survey of atomic references that could surpass the two-level lifetime limit, as well as a detailed analysis of a clock experiment using a single $^{27}\text{Al}^{+}$ quantum logic clock.

\subsection{Theoretical model}

\par
\begin{figure}
    \centering
    \includegraphics[width=1.05\linewidth]{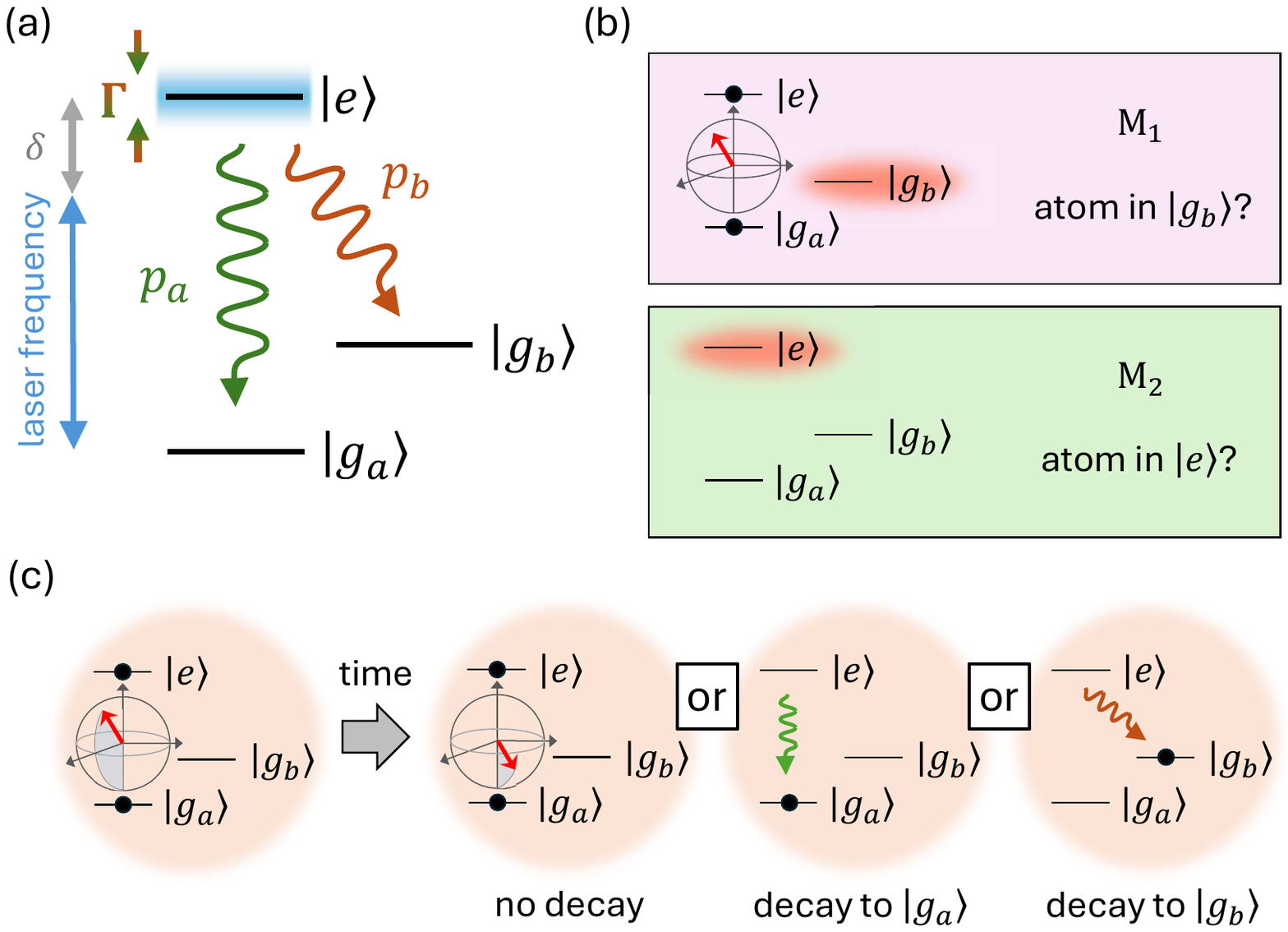}
    \caption{Three-level atom model. (a) A three level atom with excited level $\left|e\right\rangle$ having a linewidth $\Gamma$ can spontaneously decay to ground levels $\left|g_{a}\right\rangle$ and $\left|g_{b}\right\rangle$ with probabilities $p_{a}$ and $p_{b}$, respectively. (b) The model assumes two projective state measurements: $\text{M}_{1}=\left|g_{b}\right\rangle\left\langle g_{b}\right|$ reveals whether the atom is in state $\left|g_{b}\right\rangle$ without affecting other populated states, while $\text{M}_{2}=\left|e\right\rangle\left\langle e\right|$ detects if the atom is in the excited state. (c) An atom is prepared in a coherent superposition of the states $\left|g_{a}\right\rangle$ and $\left|e\right\rangle$, denoted schematically by a Bloch sphere representation, and where populated levels are depicted by black circles. At later times, the atom state is a mixed state of the three-level system: when no decay occurred, the atom is in a superposition of the states $\left|g_{a}\right\rangle$ and $\left|e\right\rangle$. This superposition acquires a phase proportional to the detuning of the laser probe from the atomic transition, and has lower excited state amplitude due to partial measurement by the atom's environment. When the atom decays, it can end up either in $\left|g_{a}\right\rangle$ or in $\left|g_{b}\right\rangle$.}
    \label{fig:theoretical model}
\end{figure}

 To show the resulting stability advantage we consider an atom with three energy levels: an excited state denoted $\left|e\right\rangle$ and two ground states denoted $\left|g_{a}\right\rangle$ and $\left|g_{b}\right\rangle$. The state $\left|e\right\rangle$ spontaneously decays at rate $\Gamma$, or equivalently with decay time $T_{d}=\frac{1}{\Gamma}$. A decay event leaves the atom in states $\left|g_{a}\right\rangle$ or $\left|g_{b}\right\rangle$ with probabilities $p_{a}$ and $p_{b}=1-p_{a}$ respectively (Fig. \ref{fig:theoretical model} a). The atomic transition frequency is interrogated using a Ramsey-type scheme (composed of state preparation pulse, wait time and closing pulse) concluding with a sequence of two projective measurements. The first, described with the observable $\text{M}_{1}=\left|g_{b}\right\rangle\left\langle g_{b}\right|$, measures whether the atom decayed to the ground state $\left|g_{b}\right\rangle$ and leaves the atomic state unperturbed otherwise. The second, described with the observable $\text{M}_{2}=\left|e\right\rangle\left\langle e\right|$, measures whether the atom is in the excited state. The concatenation of the two measurements unambiguously reveals which of the three levels the atom collapsed into. Practically, both measurements can be implemented with a combination of fluorescence detection and shelving pulses to non-fluorescing levels, or quantum logic spectroscopy (QLS).


\par We measure the performance of a frequency interrogation protocol based on its asymptotic stability at long averaging time. We denote the population outcome as $P$, which depends on the detuning $\delta$ between the laser and the frequency difference between the levels $\left|g_{a}\right\rangle$ and $\left|e\right\rangle$, the interrogation time $T$ and an experimental phase $\phi$ (typically a laser phase), among additional experimental parameters. Assuming quantum projection noise averaged over total number of interrogations $N_{\text{tot}}$, which generally depends on the total averaging time $\tau$ and the interrogation time $T$, the population measurement noise is given by 
\begin{equation} \label{standard deviation calculation}
    \sigma=\left(N_{\text{tot}}^{-1} P\left(1-P\right)\right)^{\frac{1}{2}}.
\end{equation}
Since the maximal sensitivity is obtained at the maximal slope of $P$ with respect to $\delta$, which occurs at $\delta T-\phi=\frac{\pi}{2} + n\pi$ for some integer $n$, we define the resulting sensitivity as
\begin{equation} \label{sensitivity definition}
    \text{S}=\left.\left(\frac{\frac{\partial P}{\partial\delta}}{\sigma}\right)^{2}\right|_{\delta T-\phi=\frac{\pi}{2}+n\pi}.
\end{equation}
The fractional frequency stability $\sigma_{\frac{\Delta \nu}{\nu}}$ of a clock operating at frequency $\nu$ is given by 

\begin{equation}
\label{eq:Stability}
    \sigma_{\frac{\Delta \nu}{\nu}}=\left(\nu \sqrt{S}\right)^{-1}
\end{equation}
For the purpose of finding the absolute limit, we assume negligible dead time between probes.

\subsection{Single three-level atom} 
The Ramsey sequence begins by initializing the atom in the ground state $\left|g_{a}\right\rangle$. The state is then rotated into the (possibly unbalanced) superposition $c_{g}\left|g_{a}\right\rangle+c_{e}\left|e\right\rangle$, where, without loss of generality, we define $c_{g}$ and $c_{e}$ to be real and positive, such that $c_{g}^{2}+c_{e}^{2}=1$. It is important to mention that this rotation may require a composite pulse, as there might not exist a direct transition between $\left|g_{a}\right\rangle$ and $\left|e\right\rangle$ due to, e.g. angular momentum selection rules. We can write this state in a density matrix notation as:
\begin{equation}
\rho\left(t=0\right)=\left(c_{g}\left|g_{a}\right\rangle +c_{e}\left|e\right\rangle \right)\left(c_{g}\left\langle g_{a}\right|+c_{e}\left\langle e\right|\right).
\end{equation}
We assume that the action of rotating the state addresses the transition $\left|g_{a}\right\rangle\rightarrow\left|e\right\rangle$ and has no effect on the transition $\left|g_{b}\right\rangle\rightarrow \left|e\right\rangle$. The state is then allowed to evolve for time $t$ according to the master equation:
\begin{equation} \label{eq: single atom master equation}
    \frac{\partial}{\partial t}\rho=-i\left[S_{z},\rho\right]+\frac{\Gamma}{2}\mathcal{L}_{\Lambda_{a}}\left[\rho\right]+\frac{\Gamma}{2}\mathcal{L}_{\Lambda_{b}}\left[\rho\right],
\end{equation}
where $S_{z}=-\frac{\delta}{2}\left|g_{a}\right\rangle \left\langle g_{a}\right|+\Delta\left|g_{b}\right\rangle \left\langle g_{b}\right|+\frac{\delta}{2}\left|e\right\rangle \left\langle e\right|$ is the levels' frequency operator in the rotating frame of a laser detuned by $\delta$ from the probed transition $\left|g_{a}\right\rangle\leftrightarrow \left|e\right\rangle$. Explicitly, $\delta=\omega_{e}-\omega_{g_{a}}-\omega_{L}$ and $\Delta=\omega_{g_{b}}-\omega_{g_{a}}-\frac{\delta}{2}$, where $\omega_{g_{a}},\ \omega_{g_{b}}$ and $\omega_{e}$ are the frequencies of the levels $\left|g_{a}\right\rangle,\ \left|g_{b}\right\rangle$ and $\left|e\right\rangle$ respectively, and $\omega_{L}$ is the frequency of the probe laser. Here, $\Lambda_{i}=\left|g_{i}\right\rangle\left\langle e\right|$ is a spontaneous decay operator from $\left|e\right\rangle$ to $\left|g_{i}\right\rangle$ and $\mathcal{L}_{A}\left[\rho\right]=2A\rho A^{\dagger} - \left(A^{\dagger}A\rho+\rho A^{\dagger}A\right)$. The first term on the right-hand side of Eq. \ref{eq: single atom master equation} is the coherent evolution of the state. Within $\mathcal{L}_{A}\left[\rho\right]$ the first term describes spontaneous decay and the last term accounts for the partial measurement of the atom's state by the environment when no decay occurred. The solution is given by:
\begin{equation} \label{eq:  solution to Lindblad equation}
    \begin{split}
&\rho\left(\delta,t\right)=P_{\psi}\left(c_{e},t\right)\left|\psi\left(\delta,t\right)\right\rangle \left\langle \psi\left(\delta,t\right)\right|\\
&+P_{g_{a}}\left(c_{e},t\right)\left|g_{a}\right\rangle \left\langle g_{a}\right|+P_{g_{b}}\left(c_{e},t\right)\left|g_{b}\right\rangle \left\langle g_{b}\right|,
    \end{split}
\end{equation}
and is written as an incoherent mixture of three pure states (Fig. \ref{fig:theoretical model} c) where $P_{\psi}=c_{g}^{2}+c_{e}^{2}e^{-\Gamma t}$, $P_{g_{b}}=p_{b}c_{e}^{2}\left(1-e^{-\Gamma t}\right)$, $P_{g_{a}}=p_{a}c_{e}^{2}\left(1-e^{-\Gamma t}\right)$  and
$\left|\psi\left(\delta,t\right)\right\rangle =\frac{1}{\sqrt{P_{\psi}}}\left(c_{g}e^{i\frac{\delta}{2}t}\left|g_{a}\right\rangle +c_{e}e^{-i\frac{\delta}{2}t}e^{-\frac{\Gamma}{2}t}\left|e\right\rangle\right)$. 
The first two terms account for interrogations that resulted in the atom being within the subspace $\{\left|g_{a}\right\rangle,\left|e\right\rangle\}$, which we refer to as being successful. Here, only the first term corresponds to evolution with no spontaneous decay. However, both the first and second term, the latter being the consequence of spontaneous decay, leave the atom in the subspace $\{\left|g_{a}\right\rangle,\left|e\right\rangle\}$, therefore, both terms are considered as representing successful interrogations. The last term accounts for unsuccessful interrogations, in which the atom decayed to the state $\left|g_{b}\right\rangle$, which can be identified by the measurement $\text{M}_{1}$. Starting from a superposition between $\left|g_{a}\right\rangle$ and $\left|e\right\rangle$, even when no decay occurs, the amplitude of the state $\left|e\right\rangle$ decreases with time during the interrogation as it becomes more likely that the atom is at a lower energy level. The next step in the protocol is the closing Ramsey (composite) pulse at the interrogation time $T$, acting as $\left|g_{a}\right\rangle\rightarrow\frac{1}{\sqrt{2}}\left(\left|g_{a}\right\rangle+e^{-i\phi}\left|e\right\rangle\right)$, $\left|g_{b}\right\rangle\rightarrow\left|g_{b}\right\rangle$ and $\left|e\right\rangle\rightarrow\frac{1}{\sqrt{2}}\left(\left|e\right\rangle-e^{i\phi}\left|g_{a}\right\rangle\right)$, for some pulse phase $\phi$ (see supplemental material). 

\par After the closing Ramsey pulse, a detection sequence consisting of measurement $\text{M}_{1}$ followed by measurement $\text{M}_{2}$ is implemented, which determines the state population in all three states. A positive result for the $\text{M}_{1}$ measurement directly reveals that the atom had decayed to $\left|g_{b}\right\rangle$ during the interrogation, and (according to Eq. \ref{eq:  solution to Lindblad equation}) occurs with probability $P_{g_{b}}\left(c_{e},T\right)$. If it indeed decayed, the interrogation is deemed unsuccessful as any phase information is lost and no result is added to the frequency measurement average. Otherwise, the interrogation is successful and the outcome of the second state detection is averaged. The probability for a successful interrogation is $1-P_{g_{b}}\left(c_{e},T\right)$, making the number of successful interrogations:
\begin{equation} \label{eq: number of meas after state detection}
    \begin{split}
        &N\left(c_{e},T\right)=N_{\text{tot}}\left(1-P_{g_{b}}\left(c_{e},T\right)\right)=\\
        &\frac{\tau}{T}\left(1-p_{b}c_{e}^{2}\left(1-e^{-\Gamma T}\right)\right).
    \end{split}
\end{equation}
We note that this number depends on the amplitude $c_{e}$ and the probability $p_{b}$ that affect the probability of the atom to spontaneously decay to the state $\left|g_{b}\right\rangle$. We can calculate the probability of measuring the atom in $\left|e\right\rangle$, conditioned on the atom not being in $\left|g_{b}\right\rangle$ (see supplementary material):
\begin{equation} \label{eq: population in excited}
    P_{e|\left(\text{not }g_{b}\right)}\left(c_{e},\delta,T,\phi\right)=\frac{1}{2}+\frac{c_{e}c_{g}e^{-\frac{\Gamma}{2}t}\cos\left(\delta t-\phi\right)}{1-p_{b}c_{e}^{2}\left(1-e^{-\Gamma t}\right)}.
\end{equation} 
Using Eqs. \ref{eq: number of meas after state detection} and \ref{eq: population in excited}, we calculate the measurement standard deviation by replacing $N_{\text{tot}}$ with $N$ and inserting $P_{e|\left(\text{not }g_{b}\right)}\left(c_{e},\delta,T,\phi\right)$ into Eq. \ref{standard deviation calculation}, and use it to write the sensitivity corresponding to Eq. \ref{sensitivity definition}:
\begin{equation}\label{eq: single_atom_sensitivity_p_b}
    S\left(c_{e},\frac{T}{T_{d}}\right)=4\tau T_{d}\frac{\frac{T}{T_{d}}c_{e}^{2}c_{g}^{2}e^{-\frac{T}{T_{d}}}}{1-p_{b}c_{e}^{2}\left(1-e^{-\frac{T}{T_{d}}}\right)},
\end{equation}
where we use the unitless interrogation time $\frac{T}{T_{d}}$. 
\par Maximizing the sensitivity in Eq. \ref{eq: single_atom_sensitivity_p_b} is equivalent to optimizing the quantum Fisher information inferred from the atom with respect to $\delta$ for a given time, and therefore correspond to the upper bound on the clock stability (see supplementary material). 

\begin{figure}
    
    \includegraphics[width=1.0\linewidth]{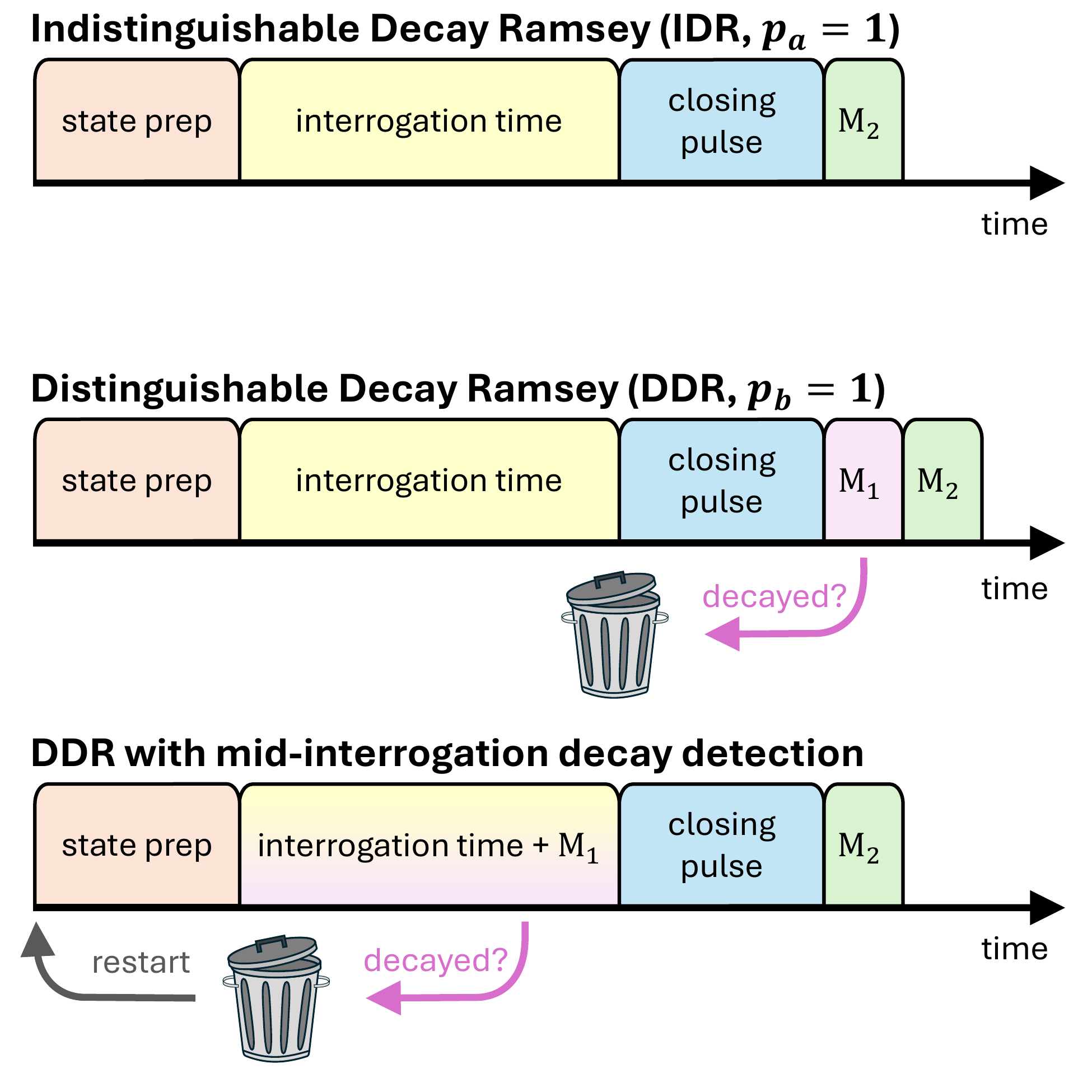}
    \caption{Interrogation protocols. All Ramsey-type protocols include a state preparation step resulting in a (possibly unbalanced) superposition, followed by an interrogation wait time that concludes with a closing pulse mapping phase into population, and finally end with a state measurement sequence. In the IDR protocol only a single $\text{M}_{2}$ measurement is applied and successful and unsuccessful interrogation results are averaged together. In the DDR protocol both $\text{M}_{1}$ and $\text{M}_{2}$ are performed at the end of the interrogation. If the $\text{M}_{1}$ decay detection is positive, the result is rejected. Otherwise, a final $\text{M}_{2}$ measurement provides the interrogation result to be averaged. In the DDR with mid-interrogation decay detection the atom is continuously checked for decay throughout the interrogation time. As soon as a decay is detected the result is discarded and the interrogation restarts. If no decay is detected by the interrogation time, a $\text{M}_{2}$ measurement produces the result to be averaged.}
    \label{fig:protocols figure}
\end{figure}

\par We analyze the protocol in two limiting cases, depicted schematically in the top and middle parts of Fig. \ref{fig:protocols figure}: the first case, which we denote indistinguishable decay Ramsey (IDR), corresponds to $p_{a}=1$, $p_{b}=0$. This case reduces to the commonly-modeled two-level atomic clock where the excited level decays to the probed transition's ground state. We denote the second case, in which $p_{a}=0$, $p_{b}=1$, as distinguishable decay Ramsey (DDR). This limit enables complete identification and exclusion of unsuccessful interrogations, which have no phase information, from the overall average, and results in improved stability compared to the IDR limit. In section \ref{sec: implementation in an ion clock} we suggest and analyze a protocol saturating the DDR bound for an $^{27}\text{Al}^{+}$ optical ion clock. 

\par In the IDR limit, the first measurement always yields a negative reading as the atom cannot decay to $\left|g_{b}\right\rangle$. Therefore, only the $\text{M}_{2}$ measurement is needed. The corresponding sensitivity takes the form
\begin{equation}
\begin{split}
    &S^{\left(\text{IDR}\right)}\left(c_{e},\frac{T}{T_{d}}\right)=4\tau T_{d}\left(\frac{T}{T_{d}}c_{e}^{2}c_{g}^{2}e^{-\frac{T}{T_{d}}}\right).
\end{split}
\end{equation}
The maximal sensitivity in this case is:
\begin{equation}
    \text{S}^{\left(\text{IDR,max}\right)}=\frac{\tau T_{d}}{e},
\end{equation}
and is obtained for the interrogation parameters $c^{\left(\text{IDR, max}\right)}_{g}=c^{\left(\text{IDR, max}\right)}_{e}=\frac{1}{\sqrt{2}}$ and $\left(\frac{T}{T_{d}}\right)^{\left(\text{IDR, max}\right)}=1$. This is the commonly used atomic clock stability limit.

\par We move to analyze the DDR limit, where $p_{b}=1$ and therefore $P_{\psi}\left(c_{e},t\right)=1-P_{g_{b}}\left(c_{e},t\right)$ result in the sensitivity
\begin{equation}
    \text{S}^{\left(\text{DDR}\right)}\left(c_{e},\frac{T}{T_{d}}\right)=4\tau T_{d}\frac{\frac{T}{T_{d}}c_{e}^{2}c_{g}^{2}e^{-\frac{T}{T_{d}}}}{c_{g}^{2}+c_{e}^{2}e^{-\frac{T}{T_{d}}}}.
\end{equation}
In this limit, the optimal balance between reduction in the number of successful interrogations and elimination of noise from unsuccessful interrogations leads to the improved sensitivity:
\begin{equation}\label{max sensitivity DDR}
    \text{S}^{\left(\text{DDR,max}\right)}\approx1.68\frac{\tau T_{d}}{ e},
\end{equation}
which is achieved with $\left(\frac{T}{T_{d}}\right)^{\left(\text{DDR, max}\right)}\approx1.48$, $c_{g}^{\left(\text{DDR, max}\right)}\approx 0.57$ and $c_{e}^{\left(\text{DDR, max}\right)}\approx 0.82$. This is an improvement of $\approx2.25\ \text{dB}$ compared to the IDR limit, and is identical to the gain mentioned in \cite{kielinski2024ghz} with multiple atoms.

It is worth noting that in the DDR case, even when the first measurement is not carried out and when averaging all interrogations, some sensitivity gain can be obtained compared to the IDR (or the two-level model) case for appropriate choice of experimental parameters. This gain arises from the fact that the second Ramsey pulse does not act on $\left|g_{b}\right\rangle$. Thus, for unsuccessful interrogations, the atom stays in $\left|g_{b}\right\rangle$ and does not end in an equal superposition of the ground and excited states that adds the maximal amount of projection noise.

\subsection{Single-excitation Bell state}
\par In the single atom superposition case, a decrease in the amplitude of state $\left|e\right\rangle$ occurs even with no spontaneous decay event. However, no information regarding the phase of the superposition leaks to the environment, and therefore superposition phase coherence is maintained. For systems with more than a single atom, it is possible to engineer a superposition state symmetric to spontaneous decay, for which both the relative amplitude and the phase between the superposition state  are insensitive to the rate of spontaneous decay. An example of such a state is a single-excitation Bell state $\left|\psi^{\left(\text{B}\right)}\left(\chi\right)\right\rangle=\frac{1}{\sqrt{2}}\left(e^{i\frac{\chi}{2}}\left|g_{a}^{\left(1\right)}e^{\left(2\right)}\right\rangle +e^{-i\frac{\chi}{2}}\left|e^{\left(1\right)}g_{a}^{\left(2\right)}\right\rangle \right)$, where $\left(i\right)$ in $g_{a}^{\left(i\right)}$ and $e^{\left(i\right)}$ indicate the $i$'th atom and $\chi$ is the superposition phase. Such states are particularly useful for atomic frequency comparisons \cite{shaniv2019quadrupole,manovitz2019precision,clements2020lifetime,zhiqiang2023176lu+} because they are insensitive to common-mode phase noise. As phase noise on the laser is typically the limiting factor in Ramsey experiments, this insensitivity to common-mode phase noise can be leveraged using correlation-spectroscopy to improve the sensitivity. 

\par In the following, we indicate properties for Bell state protocols with $\left(\text{B}\right)$. We set the starting point of the protocol with the pure state 
\begin{equation}
\begin{split}
    \rho^{\left(\text{B}\right)}\left(t=0\right)=\left|\psi^{\left(\text{B}\right)}\left(0\right)\right\rangle\Big\langle\psi^{\left(\text{B}\right)}\left(0\right)\Big|.
\end{split}
\end{equation}
This state propagates according to the master equation:
\begin{equation}
\begin{split}
    &\frac{\partial}{\partial t}\rho^{\left(\text{B}\right)}=-i\left[S_{z}^{\left(1\right)}\otimes I^{\left(2\right)}+I^{\left(1\right)}\otimes S_{z}^{\left(2\right)},\rho^{\left(\text{B}\right)}\right]+\\
    &\frac{\Gamma}{2}\mathcal{L}_{\Lambda_{a}^{\left(1\right)}\otimes I^{\left(2\right)}}\left[\rho^{\left(\text{B}\right)}\right]+\frac{\Gamma}{2}\mathcal{L}_{I^{\left(1\right)}\otimes\Lambda_{a}^{\left(2\right)}}\left[\rho^{\left(\text{B}\right)}\right]+\\
    &\frac{\Gamma}{2}\mathcal{L}_{\Lambda_{b}^{\left(1\right)}\otimes I^{\left(2\right)}}\left[\rho^{\left(\text{B}\right)}\right]+\frac{\Gamma}{2}\mathcal{L}_{I^{\left(1\right)}\otimes\Lambda_{b}^{\left(2\right)}}\left[\rho^{\left(\text{B}\right)}\right],   
\end{split}
\end{equation}
where $S_{z}^{\left(i\right)}=-\frac{\delta^{\left(i\right)}}{2}\left|g_{a}^{\left(i\right)}\right\rangle \left\langle g_{a}^{\left(i\right)}\right|+\Delta^{\left(i\right)}\left|g_{b}^{\left(i\right)}\right\rangle \left\langle g_{b}^{\left(i\right)}\right|+\frac{\delta^{\left(i\right)}}{2}\left|e^{\left(i\right)}\right\rangle \left\langle e^{\left(i\right)}\right|$ is the level frequency operator defined as in Eq. \ref{eq: single atom master equation} for atom $i$,  $I^{\left(i\right)}=\left|g_{a}^{\left(i\right)}\right\rangle \left\langle g_{a}^{\left(i\right)}\right|+\left|g_{b}^{\left(i\right)}\right\rangle \left\langle g_{b}^{\left(i\right)}\right|+\Big|e^{\left(i\right)}\Big\rangle \Big\langle e^{\left(i\right)}\Big|$ is the identity operator of atom $i$ and $\Lambda_{a}^{\left(i\right)}=\left|g_{a}^{\left(i\right)}\right\rangle \Big\langle e^{\left(i\right)}\Big|$ and $\Lambda_{b}^{\left(i\right)}=\left|g_{b}^{\left(i\right)}\right\rangle \Big\langle e^{\left(i\right)}\Big|$ are the decay operators of atom $i$ to levels $g_{a}$ and $g_{b}$ respectively. 

\par The solution of the master equation is given by:
\begin{equation}\label{Bell state master equation solution}
    \begin{split}        &\rho^{\left(\text{B}\right)}\left(\delta^{\left(\text{B}\right)},t\right)=P_{\psi}^{\left(\text{B}\right)}\left(t\right)\left|\psi^{\left(\text{B}\right)}\left(\delta^{\left(\text{B}\right)}t\right)\right\rangle \left\langle \psi^{\left(\text{B}\right)}\left(\delta^{\left(\text{B}\right)}t\right)\right|\\
    &+P_{g_{a}}^{\left(\text{B}\right)}\left(t\right)\left|g_{a}^{\left(1\right)}g_{a}^{\left(2\right)}\right\rangle \left\langle g_{a}^{\left(1\right)}g_{a}^{\left(2\right)}\right|\\
    &+P_{1,g_{b}}^{\left(\text{B}\right)}\left(t\right)\left|g_{b}^{\left(1\right)}g_{a}^{\left(2\right)}\right\rangle \left\langle g_{b}^{\left(1\right)}g_{a}^{\left(2\right)}\right|\\
    &+P_{2,g_{b}}^{\left(\text{B}\right)}\left(t\right)\left|g_{a}^{\left(1\right)}g_{b}^{\left(2\right)}\right\rangle \left\langle g_{a}^{\left(1\right)}g_{b}^{\left(2\right)}\right|,
    \end{split}
\end{equation}
where $P_{\psi}^{\left(\text{B}\right)}\left(t\right)=e^{-\Gamma t}$, $P_{1,g_{b}}^{\left(\text{B}\right)}\left(t\right)=P_{2,g_{b}}^{\left(\text{B}\right)}\left(t\right)=\frac{1}{2}\left(1-e^{-\Gamma t}\right)p_{b}$, $P_{g_{a}}^{\left(\text{B}\right)}\left(t\right)=\left(1-e^{-\Gamma t}\right)p_{a}$ and where we defined $\delta^{\left(\text{B}\right)}=\delta^{\left(1\right)}-\delta^{\left(2\right)}$ as the difference in the transition frequency between the two atoms. Here, the first term describes the atom's state when no decay has occurred and the last three terms are the possible atomic states after a decay event with their corresponding probabilities. Similarly to the single atom case, the first two terms account for successful interrogations and the last two terms represent for unsuccessful ones.

\par At the interrogation time $T$ we apply a $\frac{\pi}{2}$ pulse resonant with the transition $\left|g_{a}\right\rangle\rightarrow\left|e\right\rangle$ that addresses both atoms simultaneously, followed by a state parity detection sequence. Similarly to the single atom case, this parity detection sequence is composed of two measurements. The first, defined as 

\begin{equation*}
\begin{split}
    &\text{M}_{1}^{\left(\text{B}\right)}=\left|g_{a}^{\left(1\right)}g_{b}^{\left(2\right)}\right\rangle \left\langle g_{a}^{\left(1\right)}g_{b}^{\left(2\right)}\right|+\left|g_{b}^{\left(1\right)}g_{a}^{\left(2\right)}\right\rangle \left\langle g_{b}^{\left(1\right)}g_{a}^{\left(2\right)}\right|\\
    &+\left|e^{\left(1\right)}g_{b}^{\left(2\right)}\right\rangle \left\langle e^{\left(1\right)}g_{b}^{\left(2\right)}\right|+\left|g_{b}^{\left(1\right)}e^{\left(2\right)}\right\rangle \left\langle g_{b}^{\left(1\right)}e^{\left(2\right)}\right|,
\end{split}
\end{equation*}
informs whether an atom decayed to $\left|g_{b}\right\rangle$. The second, defined as 

 \begin{equation*}
 \begin{split}
     &\text{M}_{2}^{\left(\text{B}\right)}=\Big|e^{\left(1\right)}e^{\left(2\right)}\Big\rangle \Big\langle e^{\left(1\right)}e^{\left(2\right)}\Big|+\left|g_{a}^{\left(1\right)}g_{a}^{\left(2\right)}\right\rangle \left\langle g_{a}^{\left(1\right)}g_{a}^{\left(2\right)}\right|\\
     &-\left|e^{\left(1\right)}g_{a}^{\left(2\right)}\right\rangle \left\langle e^{\left(1\right)}g_{a}^{\left(2\right)}\right|-\left|g_{a}^{\left(1\right)}e^{\left(2\right)}\right\rangle \left\langle g_{a}^{\left(1\right)}e^{\left(2\right)}\right|,
 \end{split}
\end{equation*}
 provides the parity measurement from which $\delta^{\left(B\right)}$ is inferred, where the first two terms correspond to even parity and the last two correspond to odd parity.

\par We directly analyze the protocol in the IDR and DDR limits. In the IDR limit neither atom can end up in $\left|g_{b}\right\rangle$, and therefore only the $\text{M}_{2}^{\left(\text{B}\right)}$ is needed. The probability of measuring odd parity at time $T$ is given by:
\begin{equation} \label{eq: P_B_odd_IDR}
    P^{\left(\text{B, IDR}\right)}_{\text{odd}}\left(\delta^{\left(\text{B}\right)},T,\phi\right)=\frac{1}{2}-\frac{1}{2}e^{-\Gamma T}\cos\left(\delta^{\left(\text{B}\right)} T-\phi\right), 
\end{equation}
where $\phi$ is the phase imprinted by the phase difference between the pulses addressing each atom. Using Eq. \ref{eq: P_B_odd_IDR} and Eq. \ref{standard deviation calculation} with $N_{\text{tot}}=\frac{\tau}{T}$ we calculate the standard deviation and use it to obtain the sensitivity, according to Eq. \ref{sensitivity definition}:
\begin{equation}
     \begin{split}
        \text{S}^{\left(\text{B, IDR}\right)}\left(\frac{T}{T_{d}}\right)=\tau T_{d}\left(\frac{T}{T_{d}}e^{-2\frac{T}{T_{d}}}\right),
     \end{split}
\end{equation}
where again we used the unitless interrogation time $\frac{T}{T_{d}}$. The maximal sensitivity in this case is:
\begin{equation}
     \begin{split}
        \text{S}^{\left(\text{B, IDR, max}\right)}=\frac{\tau T_{d}}{2e},
     \end{split}
\end{equation}
for $\left(\frac{T}{T_{d}}\right)^{\left(\text{B, IDR, max}\right)}=\frac{1}{2}$, which accounts for the faster spontaneous decay rate the two-atom superposition.

\par In the DDR limit any positive outcome from the $\text{M}_{1}^{\left(\text{B}\right)}$ measurement means that an atom decayed to $\left|g_{b}\right\rangle$, and therefore indicates an unsuccessful interrogation that is then discarded. According to Eq. \ref{Bell state master equation solution} the probability to have a successful interrogation at time $T$ is $P^{\left(\text{B}\right)}_{\psi}=e^{-\Gamma T}$. The probability of measuring odd parity in the $\text{M}_{2}^{\left(\text{B}\right)}$ measurement conditioned on none of the atoms being in $\left|g_{b}\right\rangle$ reads:
\begin{equation}\label{eq: P_B_DDR}
    \begin{split}
        P_{\text{ odd}|\left(\text{not }g_{b}\right)}^{\left(\text{B, DDR}\right)}\left(\delta^{\left(\text{B}\right)},T,\phi\right)=\frac{1}{2}-\frac{1}{2}\cos\left(\delta^{\left(\text{B}\right)} T-\phi\right).
    \end{split}
\end{equation}
To calculate the standard deviation, we must account only for the number of successful interrogations, which amounts to 

\begin{equation}\label{eq: N_B_DDR}
    N^{\left(\text{B, DDR}\right)}\left(T\right)=N_{\text{tot}}P^{\left(\text{B}\right)}_{\psi}\left(T\right)=\frac{\tau}{T}e^{-\Gamma T}.    
\end{equation}

 Using Eq. \ref{eq: P_B_odd_IDR} and Eq. \ref{standard deviation calculation} with $N^{\left(\text{B, DDR}\right)}\left(T\right)$ we calculate the standard deviation and use it to obtain the sensitivity:

\begin{equation}\label{Bell state DDR sensitivity}
    \begin{split}
        &\text{S}^{\left(\text{B, DDR}\right)}\left(\frac{T}{T_{d}}\right)=\tau T_{d}\left(\frac{T}{T_{d}}e^{-\frac{T}{T_{d}}}\right).
    \end{split}
\end{equation}
The maximal sensitivity is 
\begin{equation}
    \text{S}^{\left(\text{B, DDR, max}\right)}=\frac{\tau T_{d}}{e},
\end{equation}
corresponding to $\left(\frac{T}{T_{d}}\right)^{\left(\text{B, DDR, max}\right)}=1$. Here, we obtain an enhancement of $\frac{\text{S}^{\left(\text{B, DDR, max}\right)}}{\text{S}^{\left(\text{B, IDR, max}\right)}}=3\  \text{dB}$ compared with the IDR limit.

\subsection{Mid-interrogation decay detection \label{sec:decaydetect}}
\par Additional sensitivity enhancement can be obtained by incorporating mid-interrogation decay detection. This can be done by performing $\text{M}_{1}$ measurements within the interrogation time interval, which consequently allows for early detection and exclusion of an unsuccessful interrogation followed by the immediate launch of the next interrogation. As a result, more interrogations are performed within the total averaging time $\tau$ reducing the average variance. The maximal gain limit for incorporating mid-interrogation decay detections is attained when decay measurements are implemented continuously, and as soon as decay occurs the atoms' state is reset and the succeeding interrogation is launched. The interrogation sequence is schematically depicted at the bottom of Fig.~\ref{fig:protocols figure}.  

\par We calculate the sensitivity in the single atom case with continuous decay detection as follows:
the probability of the atom, initialized at time $t=0$, to decay to $\left|g_{b}\right\rangle$ between the time $t'$ and the time $t'+dt'$ is $dP_{g_{b}}=\left|c_{e}\left(t'\right)\right|^{2}\Gamma dt'$, with $\left|c_{e}\left(t'\right)\right|^{2}$ being the population in the excited state at time $t'$. According to Eq. \ref{eq:  solution to Lindblad equation} this probability is:
\begin{equation}\label{eq: dP_gb}
    dP_{g_{b}}\left(c_{e},t'\right)=p_bc^{2}_{e}e^{-\Gamma t'}\Gamma dt'.
\end{equation}

We can use this probability density along with the probability for successful interrogation $1-P_{g_b}(c_e,t)$ to determine the mean interrogation time (successful and unsuccessful):
\begin{equation}
    \begin{split}
    &T_{\text{mean}}=\int_{0}^{T} p_b c^{2}_{e}e^{-\Gamma t'}\Gamma t' dt' + \left(1-p_bc_{e}^{2}(1-e^{-\Gamma T})\right)T=\\
    &T\left(1-p_b c_{e}^{2}+\frac{p_bc_{e}^{2}}{\Gamma T}\left(1-e^{-\Gamma T}\right)\right),
    \end{split}
\end{equation}
and obtain the mean number of successful interrogation within total averaging time $\tau$:
\begin{equation}
N^{\left(\text{mid}\right)}\left(c_{e},T\right)=\frac{\tau}{T_{\text{mean}}}\left(1-P_{g_b}(c_e,t)\right)
\end{equation}
Evaluating this in the DDR limit, we get
\begin{equation}
N^{\left(\text{DDR,mid}\right)}\left(c_{e},T\right)=\frac{\tau\left(c_{g}^{2}+c_{e}^{2}e^{-\Gamma T}\right)}{T\left(c_{g}^{2}+\frac{c_{e}^{2}}{\Gamma T}\left(1-e^{-\Gamma T}\right)\right)}.
\end{equation}
We can now compute the sensitivity in this case using $N^{\left(\text{DDR, mid}\right)}$ as the number of measurements:
\begin{equation}\label{eq: S_DDR_mid}
    \begin{split}
        &\text{S}^{\left(\text{DDR, mid}\right)}\left(c_{e},\frac{T}{T_{d}}\right)=\\
        &4\tau T_{d}\frac{\left(\frac{T}{T_{d}}\right)^{2}c_{e}^{2}c_{g}^{2}e^{-\frac{T}{T_{d}}}}{\left(\frac{T}{T_{d}}c_{g}^{2}+c_{e}^{2}\left(1-e^{-\frac{T}{T_{d}}}\right)\right)\left(c_{g}^{2}+c_{e}^{2}e^{-\frac{T}{T_{d}}}\right)}.
    \end{split}
\end{equation}

On optimizing the experimental parameters, the maximal sensitivity is obtained for $c_{e}^{\left(\text{DDR, mid, max}\right)}\approx0.91$, $c_{g}^{\left(\text{DDR, mid, max}\right)}\approx0.41$ and $\left(\frac{T}{T_{d}}\right)^{\left(\text{DDR, mid, max}\right)}\approx 2.25$ and reads
\begin{equation}
\label{eq: S_DDR_mid_max}
    \begin{split}
        &\text{S}^{\left(\text{DDR, mid, max}\right)}\approx2.82\frac{\tau T_{d}}{e},
    \end{split}
\end{equation}
which is an improvement of $\frac{\text{S}^{\left(\text{DDR, mid, max}\right)}}{\text{S}^{\left(\text{IDR}\right)}}\approx4.5\text{ dB}$. This sets a new lifetime limit for frequency interrogation of uncorrelated atoms.

Next, we evaluate the performance of the Bell state case in the DDR limit with mid-interrogation decay detection. According to Eq. \ref{Bell state master equation solution}, the probability of any of the two atoms to decay to $\left|g_{b}\right\rangle$ between times $t'$ and $t'+dt'$ is 
\begin{equation}
    dP^{\left(\text{B}\right)}_{g_{b}}\left(t'\right)=e^{-\Gamma t'}\Gamma dt', 
\end{equation}
and the probability to have a successful interrogation with time $T$ is
\begin{equation}
    P^{\left(\text{B}\right)}_{\psi}\left(T\right)=e^{-\Gamma T}.
\end{equation}
The mean interrogation time then reads:
\begin{equation}
\begin{split}
    &T_{\text{mean}}^{\left(\text{B}\right)}=\\
    &\int_{0}^{T}e^{-\Gamma t'}\Gamma t'dt'+e^{-\Gamma T}T=\frac{1}{\Gamma}\left(1-e^{-\Gamma T}\right),
\end{split}
\end{equation}
and the number of successful interrogation becomes:
\begin{equation}
\begin{split}
    &N^{\left(\text{B, DDR, mid}\right)}\left(T\right)=\frac{\tau}{T_{\text{mean}}^{\left(\text{B}\right)}}P^{\left(\text{B}\right)}_{\psi}\left(T\right)=\frac{\tau e^{-\Gamma T}}{\frac{1}{\Gamma}\left(1-e^{-\Gamma T}\right)}.
\end{split}
\end{equation}

\noindent
\begin{table*}[t]
\begin{tabular}{|c||*{5}{c|}}\hline
\backslashbox{parameters}{protocol}
&\makebox[10em]{IDR}&\makebox[10em]{DDR}&\makebox[10em]{DDR mid single}
&\makebox[10em]{DDR mid}\\\hline\hline
$\frac{T}{T_{d}}=1$, $c_{g}=c_{e}=\frac{1}{\sqrt{2}}$ & $1$ & $1.46$ &$1.62^{**}$  & $1.79$ \\\hline
optimal parameters & $1$ & $1.68$ &$2.1$  & $2.82$\\\hline
\end{tabular}
\caption{Sensitivity gain comparison between different protocols with a single atom, both with optimal parameters and with IDR optimal parameters. Sensitivity gain is with respect to optimal sensitivity in the DDR case $\text{S}^{\left(\text{IDR, max}\right)}=\frac{\tau T_{d}}{e}$. Here ``mid'' refers to a continuous mid-interrogation decay detection and ``mid single'' refers to mid-interrogation decay detection at a specific time. ** Mid-interrogation detection time $\frac{1}{2}T_{d}$.}\label{table: sensitivities of single atom}
\end{table*}

\noindent
\begin{table*}[t]
\begin{tabular}{|c||*{5}{c|}}\hline
\backslashbox{parameters}{protocol}
&\makebox[10em]{IDR (B)}&\makebox[10em]{DDR (B)}&\makebox[10em]{DDR (B) mid single}
&\makebox[10em]{DDR (B) mid}\\\hline\hline
$\frac{T}{T_{d}}=\frac{1}{2}$ & $1$ & $1.64$ &$1.85^{**}$  & $2.10$ \\\hline
optimal parameters & $1$ & $2.00$ &$2.55$  & $3.52$\\\hline
\end{tabular}
\caption{Sensitivity gain comparison between different protocols with a single excitation Bell state, both with optimal parameters and with IDR optimal parameters. Sensitivity gain is with respect to optimal sensitivity in the (B, IDR) case $\text{S}^{\left(\text{B, IDR, max}\right)}=\frac{\tau T_{d}}{2e}$. Here ``mid'' refers to a continuous mid-interrogation decay detection and ``mid single'' refers to mid-interrogation decay detection at a specific time. ** Mid-interrogation detection time $\frac{1}{4}T_{d}$.}\label{table: sensitivities of bell state}
\end{table*}

We can calculate the sensitivity in this case similarly to Eq. \ref{Bell state DDR sensitivity}, but taking $N^{\left(\text{B, DDR, mid}\right)}\left(T\right)$ as the number of measurements:

\begin{equation}
    \text{S}^{\left(\text{B, DDR, mid}\right)}\left(\frac{T}{T_{d}}\right)=\tau T_{d}\frac{\left(\frac{T}{T_{d}}\right)^{2}e^{-\frac{T}{T_{d}}}}{1-e^{-\frac{T}{T_{d}}}}.
\end{equation}

The maximal value is obtained for $\left(\frac{T}{T_{d}}\right)^{\left(\text{B, DDR, mid, max}\right)}\approx1.6$ and amounts to

\begin{equation}
    \text{S}^{\left(\text{B, DDR, mid, max}\right)}\approx3.5\frac{\tau T_{d}}{2 e},
\end{equation}
which is a gain of $\frac{\text{S}^{\left(\text{B, DDR, mid, max}\right)}}{\text{S}^{\left(\text{B, IDR, max}\right)}}\approx5.4\text{ dB}$.

Continuous mid-interrogation measurements provide a new stability lifetime limit. However, performing decay detection continuously may be technically challenging to implement. A more practical approach would be to perform a single mid-interrogation decay detection at a specific time within the interrogation interval. Simultaneous optimization of the mid-measurement time and the total interrogation time (and for the single atom case the initial rotation amplitudes $c_{g}$ and $c_{e}$) results in a gain of $\approx3.2 \text{ dB}$ for the single atom case and a gain of $4.0\text{ dB}$ for the Bell state case (see supplementary material). We summarize the sensitivities of the protocols analyzed in this work in tables \ref{table: sensitivities of single atom} and \ref{table: sensitivities of bell state}, where in addition to the optimal sensitivities, we show that sensitivity enhancement emerges even when keeping the interrogation parameters at the two-level model optimized parameters ($c_{g}=c_{e}=\frac{1}{\sqrt{2}}$ and $\frac{T}{T_{d}}=1$ for the single atom case and $\frac{T}{T_{d}}=\frac{1}{2}$ for the Bell state case).

\subsection{Implementation in trapped ion optical clocks}\label{sec: implementation in an ion clock}

\begin{table*}[t!]
\begin{center}
\begin{tabular}{c c c S[table-format=3.1] S[table-format=4.1] S[table-format=2.6] c}
\hline
\hline
\\[-3mm]
Ion & I & Transition & \text{$\lambda$ [nm]} & \text{$\nu$ [THz]} & \text{$\tau_e$} [s] & $\sigma_{\Delta f/f}^{\left(\text{DDR, mid}\right)}(1\,{\rm s})$ \\ \hline
\\[-2.5mm]
 $^{27}$Al$^+$~\cite{rosenband2007observation} & 5/2 & $\rm {^1S_0}-{^3P_0}$ & 267.4 & 1121.0 & 20.6(1.4) & 3.1e-17\\
 $^{40}$Ca$^+$~\cite{kreuter2005experimental} & 0 & $\rm {^2S_{1/2}}-{^2D_{5/2}}$ & 729.3 & 411.0 & 1.168(9) & 3.5e-16\\
 $^{40}$Ar$^{13+}$~\cite{king2022optical} & 0 & $\rm {^2P_{1/2}}-{^2P_{3/2}}$ & 441.4 & 679.2 & 0.009573(6) & 2.4e-15\\
 $^{88}$Sr$^+$~\cite{dube2015sr} & 0 & $\rm {^2S_{1/2}}-{^2D_{5/2}}$ & 674.0 & 444.7 & 0.3908(16) & 5.6e-16\\
 $^{115}$In$^+$~\cite{hausser2025in+} & 9/2 & $\rm {^1S_0}-{^3P_0}$ & 236.5 & 1267.4 & 0.195(8) & 2.8e-16\\
 $^{171}$Yb$^+$~\cite{tamm2014cs} & 1/2 & $\rm {^2S_{1/2}}-{^2D_{3/2}}$ & 435.5 & 688.4 & 0.0527(24) & 9.9e-16\\
 $^{199}$Hg$^{+}$~\cite{rafac2000sub} & 1/2 & $\rm {^2S_{1/2}}-{^2D_{5/2}}$ & 281.6 & 1064.7 & 0.090(1)& 4.9e-16\\
\hline
\end{tabular}
\caption{\label{tab:state_props} Lifetime-limited stability for selected single-ion clocks calculated using Eqs.~\ref{eq:Stability} and \ref{eq: S_DDR_mid_max}.  The data include transition specifications and properties for several systems that have been demonstrated experimentally as optical frequency standards. Wavelengths and frequencies are truncated at 4 digits, with more precise values available in the literature (e.g. ~\cite{ludlow2015optical}).  This table excludes other candidates such as $^{171}$Yb$^+$ (octupole transition) and $^{175}$Lu$^+$, which have excited state lifetimes many orders of magnitude beyond the coherence times of existing laser technology and thus are unlikely to reach their respective lifetime limits.}
\end{center}
\end{table*}

In Tab.~\ref{tab:state_props} we list several of the most common ion clock species and the atomic properties most relevant for their lifetime-limited stability. 
In what follows we provide an experimental protocol for the particular case of $\Al$ and consider issues such as coherence time and systematic shifts.  

\begin{figure}[hbt!]
\begin{centering}
\includegraphics[width=1.0\linewidth]{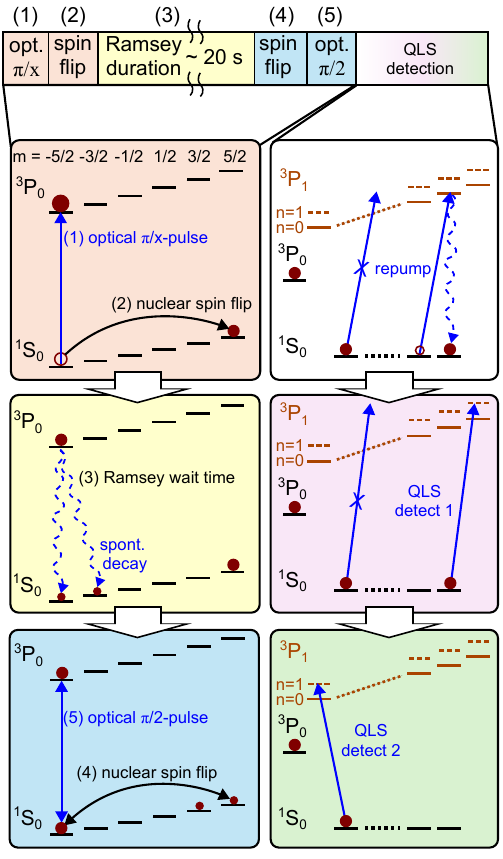}
\caption{\label{fig:AlDDR} $^{27}$Al$^+$ Distinguishable Decay Ramsey (DDR) protocol. The steps depicted in the timing diagram (top) correspond to the general procedure shown in Fig.~\ref{fig:protocols figure}. We propose a Ramsey sequence that produces a superposition $c_g\ket{\S, m=+5/2} + c_e\ket{\P, m=-5/2}$ by combining an initial optical $\pi/x$ pulse with $x\le2$ and a nuclear spin flip pulse. These steps are shown in the energy level diagrams on the left column.  To close the Ramsey sequence, two pulses reverse the sequence above using a $\pi/2$ optical pulse, which is optimal for all protocols discussed here. The right column shows the $\Al$ laser pulses involved in a quantum logic spectroscopy (QLS) procedure, which is designed to distinguish population in states $\ket{\P, m= +5/2}$, $\ket{\S, m=-5/2}$ and $\{\ket{\S, m=+3/2}, \ket{\S, +5/2}\}$ corresponding respectively to $\ket{e}$, $\ket{g_a}$ and $\ket{g_b}$ in the general protocol. This process involves excitation of the ion motion from the ground state ($n=0$) to an excited Fock state ($n=1$), dependent on the internal state of the ion~\cite{schmidt2005spectroscopy}. The motional state is then detected using sideband transitions on the qubit ion followed by resonance fluorescence detection. Note that steps in the process involving laser pulses on the qubit ion are omitted here.}
\end{centering}
\end{figure}

The $\Al$ optical clock is based on the $\S\leftrightarrow{\P}$ transition at 267.4 nm ($1.121\times10^{15}$~Hz).  The nuclear spin $I=5/2$ produces six Zeeman sublevels $m = \{-5/2, -3/2,...5/2\}$ in both the ground and excited states.  Optical clocks based on this transition typically average the frequency of two transitions on opposite sides of these manifolds to eliminate the first-order Zeeman shift.  Because $\Al$ does not have an easily accessible cycling transition suitable for laser cooling and fluorescence state detection, all clocks based on this transition have been developed using a second ion species for sympathetic cooling and state detection, which is known as quantum logic spectroscopy (QLS). Qubit species $^9{\rm Be}^+$, $^{25}{\rm Mg}^+$, and $^{40}{\rm Ca}^+$ have been explored experimentally for this role. More details can be found in the review paper~\cite{ludlow2015optical} and references therein.  

Fig.~\ref{fig:AlDDR} shows one possible experimental sequence that implements the DDR scheme in $\Al$.  It depicts a generalized Ramsey experiment, in which the superposition of states during the Ramsey wait time is composed of two Zeeman sublevels on opposite sides of the $\S$  and $\P$ manifolds, respectively. This ensures that the spontaneous decay channel from the $\P$ state is outside of the subspace of the Ramsey superposition. It can be accomplished by introducing a resonant radio-frequency (RF) spin-flip ($\pi$-pulse), which maps population according to $\ket{\S, -m}\leftrightarrow\ket{\S, +m}$. After the Ramsey wait time, the sequence is reversed, using a second RF spin flip and a $\pi/2$-pulse on the optical transition. Finally, the state populations are detected.  To filter out spontaneous decay events, the detection must distinguish population in the subspace $\{\ket{\S, m = +5/2}, \ket{\S, m = +3/2}\}$ in addition to the clock states $\ket{\P, m = -5/2}$ and $\ket{\S, m = -5/2}$. This can be done through a multi-step QLS protocol that sequentially maps state populations to the qubit excitation of a ``logic'' ion which can be directly detected~\cite{schmidt2005spectroscopy}.  In adding these steps to the experimental sequence, several potential issues arise including coherence times, systematic shifts, and state detection fidelity, which are addressed individually below. 

\subsubsection{Coherence time}

The wavelength of the $\SP$ transition in $\Al$ is 267.4~nm  with an excited state lifetime of 20.6(1.4)~s ~\cite{rosenband2007observation}. Current state-of-the-art laser coherence times~\cite{lee2026frequency} are insufficient to observe a lifetime-limited stability for this transition, but an improvement in laser coherence by a factor of 10, achievable at the thermal noise limit of cryogenic optical cavities under development~\cite{valencia2024cryogenic}, would be sufficient to achieve this in the near future. On the other hand, correlation spectroscopy experiments on the $\SP$ transition have already observed contrast and stability consistent with the lifetime limit, indicating potential for significant improvement through the DDR techniques described here.

The differential magnetic field sensitivity of the clock states $\ket{\S, m=+5/2}$ and $\ket{\P, m=-5/2}$ is 96.9~kHz/mT. Maintaining atomic coherence during a probe time of 20 s or greater requires magnetic field stability better than 0.1~$\rm{nT}$. Passive magnetic shielding in cryogenic~\cite{leopold2019cryogenic} or room temperature~\cite{ruster2016long} environments may be able to achieve this, but other techniques can reduce the requirements substantially. First, the use of non-extreme Zeeman levels, such as $\ket{\S, m=+3/2}$ and $\ket{\P, m=-1/2}$, for the clock states  reduces the sensitivity by more than a factor of 3 while maintaining the condition that spontaneous decay lies strictly outside of the Ramsey subspace. A superposition of these states can be prepared in the same way described above except initializing the ion(s) in $\ket{\S, m=-3/2}$~\cite{clements2020lifetime}. Alternatively, the states $\ket{\S, m=-1/2}$ and $\ket{\P, m=-1/2}$ exhibit a differential magnetic field sensitivity 12 times lower than the nominal 96.9~kHz/mT and a branching ratio outside of the Ramsey superposition greater than 97~\%. Another approach to reduce sensitivity to low frequency magnetic field noise is dynamical decoupling using one or more RF spin flips on both the excited state and ground state during the Ramsey wait time~\cite{akerman2025operating}.  

\subsubsection{Systematic shifts}
The addition of RF spin-flips as part of the Ramsey sequence introduces AC Zeeman shifts through two distinct mechanisms.  First, the RMS magnetic field during the spin-flip couples the $\P$ and $\Pone$ levels leading to a shift on the clock transition at the level of -71.944(24)~mHz/mT$^2$~\cite{brewer2019measurements}.  For a 20~s Ramsey wait time, we take a nominal 0.1~s rf $\pi$-pulse duration, corresponding to a peak magnetic field of 600 nT assuming an rf magnetic field with linear polarization perpendicular to the quantization field. This field amplitude is less than 1~\% of the strength of the quantization field and is applied for 1~\% of the duty cycle, leading to negligible modification to the quadratic Zeeman shift, even with clock uncertainty at the level of $10^{-19}$.

A second source of AC Zeeman shift during the rf spin-flip is due to off-resonant coupling of the $\P$ Zeeman levels to the field.  We can estimate this based on the g-factor ratio $g_p/g_s \sim 2.5$~\cite{rosenband2007observation} corresponding to ratio of on-resonance Rabi rates for the $\P$ state compared to the $\S$.  The detuning from resonance is given by $\hbar\delta_{RF} = (g_p-g_s)\mu_B B_{dc}$, where $\mu_B$ is the Bohr magneton and $\hbar$ is the reduced Planck constant, resulting in a shift of less than 1~mHz at a bias field $B_{dc} = 0.1$~mT. Since this is applied for only 1~\% of the Ramsey wait time, its effect on the average frequency of the clock is reduced by another factor of 100.  In addition, the sign of this shift changes depending on the side of the Zeeman structure probed, so it can be canceled through alternating probes on opposite sides of the Zeeman structure.   

\subsubsection{State detection}
State-dependent detection of the clock states can be implemented with high-fidelity using repeated QLS sequences.  In our proposed implementation, spontaneous emission events through both decay channels $\ket{\P,m=+5/2}\rightarrow\ket{\S, m=+3/2}$ and $\ket{\P,m=+5/2}\rightarrow\ket{\S, m=+5/2}$ must be detected to maximize clock stability.  This can be accomplished by first repumping $\ket{\S, m=+3/2}$ to $\ket{\S, m=+5/2}$, then performing a QLS sequence as usual on the cycling transition $\ket{\S, m=+5/2}\leftrightarrow\ket{\Pone, m=+7/2}$, as sketched in Fig.~\ref{fig:AlDDR}.  The requirement of distinguishing the clock ground state $\ket{\S, m=-5/2}$ from the two decay channels,  adds a condition that the optical pumping pulses and the QLS pulses must not depump population in $\ket{\S, m=-5/2}$.  This is dependent on the linear Zeeman splitting in the $^3\rm{P}_1$ state (6~MHz/mT) and the motional sideband frequencies.  At a typical bias field of 0.1~mT, the states $\ket{\S, m=+3/2}$ and $\ket{\S, m=+5/2}$ have resonances with the $^3\rm{P}_1$ states shifted by $\sim2.4$~MHz and $\sim3.0$~MHz, respectively, compared to those of the state $\ket{\S, m=-5/2}$. These lie within the range of motional frequencies (typically 1 to 10 MHz) leading to the possibility of accidental resonances that pump population toward the target state for detection. These can be avoided by careful choice of the motional spectrum and quantization field to avoid degeneracies.  For example, using the conditions described in Ref.~\cite{marshall2025high} (motional frequencies: $[f_x, f_y, f_z] = [4.38, 5.44, 2.19]$~MHz for a single $^{25}{\rm Mg}^+$, $B_{dc} = 0.100$~mT, and $\pi$-pulse duration $t_{\pi} = 2$~${\rm\mu s}$ on the $^3P_1$ cycling transition), we estimate a probability of depumping the state $\ket{\S, m=-5/2}$ of $P_{depump} = 6\times10^{-3}$ per repumping pulse and $P_{depump} = 7\times10^{-4}$ per sideband pulse.  To reach a target fidelity of 99~\% for this detection, we need approximately 10 repump pulses and 4 sideband pulses resulting in a total probability of depumping above 5~\%. However, if the Rabi rate on the pumping transition is reduced by a factor of 4, we estimate a total probability of depumping of 0.3~\%.  If this approach to optimizing detection fidelity proves insufficient in practice, one or more Zeeman states in $\S$ could be shelved to $\P$ before the QLS detection sequence, avoiding the issue of depumping withing the $\S$ manifold.

\subsubsection{Mid-interrogation state detection}

The addition of one or more mid-interrogation measurements that identify population in $\ket{g_b}$ could further optimize clock stability as described in Sec.~\ref{sec:decaydetect}.  This requires performing the measurement without perturbing the coherence of the clock superposition and without adding appreciable systematic shifts.  In general, this is a difficult problem because fluorescence measurements require a relatively strong interaction with a laser beam and the environment, through scattered photons. 

Here, QLS offers an advantage because the part of the measurement interaction applied directly to the clock ion only needs to produce state-dependent ion motion. Mid-interrogation state detection follows the same procedure outlined above to detect populations in the decay channels ${\ket{\S, m=-5/2}}$ and ${\ket{\S, m=-3/2}}$. Starting from the condition depicted in Fig.~\ref{fig:AlDDR}, step (3), the state ${\ket{\S, m=-3/2}}$ is first resonantly repumped to ${\ket{\S, m=-5/2}}$.  Then population in $\ket{\S, m=-5/2}$ is detected using QLS by applying a motional sideband pulse on the transition ${\ket{\S, m=-5/2}\leftrightarrow\ket{\Pone, m=-7/2}}$. 

The calculations of depumping probability in the previous section applying equally well to this case. For a mid-interrogation detection, we also need to consider the AC Stark shifts due to the repumping and sideband pulses.  To determine AC Stark shifts on state $\ket{\S, m=+5/2}$ due to the repumping pulses, we consider off-resonant carrier coupling given by $\Delta_{S} = c_{\sss (5/2, 3/2)}^2\Omega_0^2/12\omega_z$, where $c_{\sss (5/2, 3/2)}$ is the coupling coefficient of the $\ket{\S, m = 5/2} \leftrightarrow \ket{\Pone, m = 3/2}$ transition, $\Omega_0$ is the on-resonance carrier Rabi rate for the $\ket{\S, m = \pm5/2} \leftrightarrow \ket{\Pone, m = \pm7/2}$ cycling transition and $\omega_z$ is the Zeeman splitting between adjacent levels in the $\Pone$ state.  The total phase shift due to $N_{R}$ cycles of this pulse (duration $t_{pulse}$) is given by $\Delta\phi = N_{R}\Delta_{S}t_{pulse}$ and the fractional frequency shift is $\Delta\nu/\nu = \Delta\phi/(2\pi T\nu_0)$. For the same conditions above with $t_{\pi} = \pi/\Omega_0 = 8$~${\rm \mu s}$ and 10 repump cycles, the phase shift is 0.015 rad, corresponding to a $1.1\times10^{-19}$ fractional frequency shift at a probe time of 20 s.  If we also consider phase shifts due to motional sideband resonances and intercombination lines up to second order, this value increases to $1.5\times10^{-19}$.  A similar calculation can be performed for the sideband pulses during QLS detection where the laser light will be farther detuned from resonance by motional frequency $\omega_m$ but the pulse time will be increased by a factor $1/\eta$ where $\eta$ the Lamb-Dicke parameter.  For typical conditions, an average number of detections $N_{det} \sim 6$ are used to reach a high detection fidelity, resulting in a total phase shift of 0.042 rad, or a fractional frequency shift of $3\times10^{-19}$.  Because these phase shifts can be calibrated by measurements directly on the ion, we anticipate that their uncertainties can be controlled to a substantially lower level than the shifts themselves. 

Another source of Stark shift is the ground state cooling pulses applied to the qubit ion during the detection sequence. The size of this shift will depend on the qubit ion used and the details of the ground state cooling pulses. We estimate for Mg$^+$ using 100 Raman sideband pulses with total saturation parameter on the D2 line of $s=100$ that the shift will be $4\times10^{-19}$~\cite{marshall2025high}.  
Added time dilation shift due to heating during fluorescence detection is another potential systematic effect to consider.  At the expense of some detection efficiency, the fluorescence detection can be performed at the Doppler cooling frequency such that the ions remain close to the Doppler limit. 

If Stark shifts and/or time dilation shifts during mid-interrogation readout become a significant limiting factor, another option would be to shelve the population in $\S$ to a Zeeman sublevel in the $\P$ state during detection, then shelve back to continue the clock interrogation.  We note that any clock species with direct fluorescence detection would likely require this kind of temporary shelving approach to avoid decohering the clock superposition.   

\section{Conclusion}
We identify clock interrogation protocols based on a generalized Ramsey sequence that are capable of reaching a fundamental quantum limit for frequency stability, set by spontaneous decay, beyond that identified previously.  The key insight is that treating the atom as a multi-level quantum system rather than the two-level approximation in earlier work allows for the detection of spontaneous decay events and removal of the associated noise from the clock measurement. Optimizing experimental parameters within this paradigm requires adjusting amplitudes in the Ramsey superposition and the interrogation times beyond the excited state lifetime, reaching a gain of 2.25 dB in terms of frequency variance.  For a correlation spectroscopy measurement either with or without an entangled input state, the gain achieved is 3 dB.  We show that these performance gains saturate the quantum Fisher information bounds. The frequency variance can be further improved by performing mid-interrogation measurements, which detect population in the states resulting from spontaneous decay. This reduces the effective dead time and achieves an ultimate gain of 4.5 dB compared to a standard clock protocol and 5.5 dB for correlation spectroscopy. 

These improvements are relevant for any measurement dominated by projection noise at the lifetime limit of an atomic system.  They could be immediately relevant for trapped-ion optical clocks where several systems, including $^{88}{\rm Sr}^+$ and $^{115}{\rm In}^+$, are reaching a stability close to the lifetime limit.  For trapped-ion clocks with longer-lived excited states, such as $\Al$, laser coherence issues have thus far prevented systems from operating close to lifetime-limited stability.  However, correlation spectroscopy allows probe times well beyond the laser coherence times. We propose a particular realization of the protocol in the case of $\Al$ and analyze the requirements for coherence, high-fidelity state detection and systematic shifts.  We find that these issues can all be addressed with techniques that have already been demonstrated, making it possible to achieve a significant stability gain without a significant loss of clock accuracy.

During the final stages of preparation of this manuscript, we became aware of a related paper~\cite{arieli2026optimal} that obtains similar metrological bounds.  

\subsection{Acknowledgments}
We thank Raghavendra Srinivas for useful discussions of spontaneous decay in optical clocks and Lindsay Sonderhouse for careful reading of the manuscript. We acknowledge support from the National Institute of Standards and Technology (NIST), the Office of Naval Research (ONR), and 
\bibliography{stability_limit_paper_references.bib}

\end{document}


\title{Supplementary material: Extending the fundamental limit to atomic clock stability}

\maketitle
\onecolumngrid

\section{Full derivation of the single ion three-level model}

We begin from the initialized state:
\begin{equation}
\rho\left(t=0\right)=\left(c_{g}\left|g_{a}\right\rangle +c_{e}\left|e\right\rangle \right)\left(c_{g}\left\langle g_{a}\right|+c_{e}\left\langle e\right|\right),
\end{equation}
which evolves according to the master equation:
\begin{equation} \label{eq: single atom master equation}
    \frac{\partial}{\partial t}\rho=-i\left[S_{z},\rho\right]+\frac{\Gamma}{2}\mathcal{L}_{\Lambda_{a}}\left[\rho\right]+\frac{\Gamma}{2}\mathcal{L}_{\Lambda_{b}}\left[\rho\right],
\end{equation}
where $S_{z}=-\frac{\delta}{2}\left|g_{a}\right\rangle \left\langle g_{a}\right|+\Delta\left|g_{b}\right\rangle \left\langle g_{b}\right|+\frac{\delta}{2}\left|e\right\rangle \left\langle e\right|$ is the atom's frequency operator in the rotating frame of the laser, $\delta=\omega_{e}-\omega_{g_{a}}-\omega_{L}$ and $\Delta=\omega_{g_{b}}-\omega_{g_{a}}-\frac{\delta}{2}$, where $\omega_{g_{a}},\ \omega_{g_{b}}$ and $\omega_{e}$ are the frequencies of the levels $\left|g_{a}\right\rangle,\ \left|g_{b}\right\rangle$ and $\left|e\right\rangle$ respectively, and $\omega_{L}$ is the frequency of the probe laser. Here, $\Lambda_{i}=\left|g_{i}\right\rangle\left\langle e\right|$ is a spontaneous decay operator from $\left|e\right\rangle$ to $\left|g_{i}\right\rangle$ and $\mathcal{L}_{A}\left[\rho\right]=2A\rho A^{\dagger} - \left(A^{\dagger}A\rho+\rho A^{\dagger}A\right)$. The solution at time $t$ is given by:

\begin{equation} \label{eq: supple  solution to Lindblad equation}
    \begin{split}
        &\rho\left(\delta,t\right)=\\
        &\underset{\text{atom did not decay}}{\underbrace{\left(c_{g}e^{i\frac{\delta}{2}t}\left|g_{a}\right\rangle +c_{e}e^{-i\frac{\delta}{2}t}e^{-\frac{\Gamma}{2}t}\left|e\right\rangle \right)\left(c_{g}e^{-i\frac{\delta}{2}t}\left\langle g_{a}\right|+c_{e}e^{i\frac{\delta}{2}t}e^{-\frac{\Gamma}{2}t}\left\langle e\right|\right)}}+\underset{\text{atom decayed to }\left|g_{a}\right\rangle }{\underbrace{p_{a}c_{e}^{2}\left(1-e^{-\Gamma t}\right)\left|g_{a}\right\rangle \left\langle g_{a}\right|}}+\underset{\text{atom decayed to }\left|g_{b}\right\rangle }{\underbrace{p_{b}c_{e}^{2}\left(1-e^{-\Gamma t}\right)\left|g_{b}\right\rangle \left\langle g_{b}\right|}}.
    \end{split}
\end{equation}
After the closing Ramsey pulse (at some phase $\phi$) the state becomes:
\begin{equation}\label{eq: full density matrix solution with pa and pb}
     \begin{split}
         &\rho\left(\delta,t\right)=\left(\frac{1}{2}\left(c_{g}^{2}+c_{e}^{2}e^{-\Gamma t}\right)-c_{e}c_{g}e^{-\frac{\Gamma}{2}t}\cos\left(\delta t-\phi\right)\right)\left|g_{a}\right\rangle \left\langle g_{a}\right|+\bm{\frac{1}{2}p_{a}c_{e}^{2}\left(1-e^{-\Gamma t}\right)\left|g_{a}\right\rangle \left\langle g_{a}\right|}\\
         &+e^{i\phi}\left(\frac{1}{2}\left(c_{g}^{2}-c_{e}^{2}e^{-\Gamma t}\right)+ic_{g}c_{e}e^{-\frac{\Gamma}{2}t}\sin\left(\delta t-\phi\right)\right)\left|g_{a}\right\rangle \left\langle e\right|+\bm{\frac{1}{2}p_{a}c_{e}^{2}\left(1-e^{-\Gamma t}\right)e^{i\phi}\left|g_{a}\right\rangle \left\langle e\right|}\\
         &+e^{-i\phi}\left(\frac{1}{2}\left(c_{g}^{2}-c_{e}^{2}e^{-\Gamma t}\right)-ic_{g}c_{e}e^{-\frac{\Gamma}{2}t}\sin\left(\delta t-\phi\right)\right)\left|e\right\rangle \left\langle g_{a}\right|+\bm{\frac{1}{2}p_{a}c_{e}^{2}\left(1-e^{-\Gamma t}\right)e^{-i\phi}\left|e\right\rangle \left\langle g_{a}\right|}\\
         &+\left(\frac{1}{2}\left(c_{g}^{2}+c_{e}^{2}e^{-\Gamma t}\right)+c_{e}c_{g}e^{-\frac{\Gamma}{2}t}\cos\left(\delta t-\phi\right)\right)\left|e\right\rangle \left\langle e\right|+\bm{\frac{1}{2}p_{a}c_{e}^{2}\left(1-e^{-\Gamma t}\right)\left|e\right\rangle \left\langle e\right|}\\
         &+\bm{p_{b}c_{e}^{2}\left(1-e^{-\Gamma t}\right)\left|g_{b}\right\rangle \left\langle g_{b}\right|},
     \end{split}
 \end{equation}
where bold font indicates the terms associated with a spontaneous decay event. 

At this point in the protocol, an $\text{M}_{1}$ measurement is performed, revealing whether a decay event leaving the atom in $\left|g_{b}\right\rangle$ occurred, without affecting interrogations for which the atom did not decay. This measurement would detect the state $\left|g_{b}\right\rangle$ with probability $p_{b}c_{e}^{2}\left(1-e^{-\Gamma t}\right)$, and all interrogations with such a result are discarded. When $\left|g_{b}\right\rangle$ is not recorded, a following $\text{M}_{2}$ measurement is performed. According to Eq. \ref{eq: full density matrix solution with pa and pb} the probability to detect the state $\left|e\right\rangle$ in the second measurement conditioned on not detecting $\left|g_{b}\right\rangle$ in the first at interrogation time $T$ is given by:
\begin{equation}
    \begin{split}
        P_{e| \text{not }g_{b}}\left(c_{g},\delta,T,\phi\right)=\frac{\frac{1}{2}\left(c_{g}^{2}+c_{e}^{2}\left(e^{-\Gamma T}+p_{a}\left(1-e^{-\Gamma T}\right)\right)\right)+c_{e}c_{g}e^{-\frac{\Gamma}{2}T}\cos\left(\delta T-\phi\right)}{1-p_{b}c_{e}^{2}\left(1-e^{-\Gamma T}\right)}=\frac{1}{2}+\frac{c_{e}c_{g}e^{-\frac{\Gamma}{2}T}\cos\left(\delta T-\phi\right)}{1-p_{b}c_{e}^{2}\left(1-e^{-\Gamma T}\right)},
    \end{split}
\end{equation}
where we see that $p_{b}>0$ reduces the denominator and improves the probability dependence on $\delta$.

\begin{figure}
    \centering
    \includegraphics[width=0.9\linewidth]{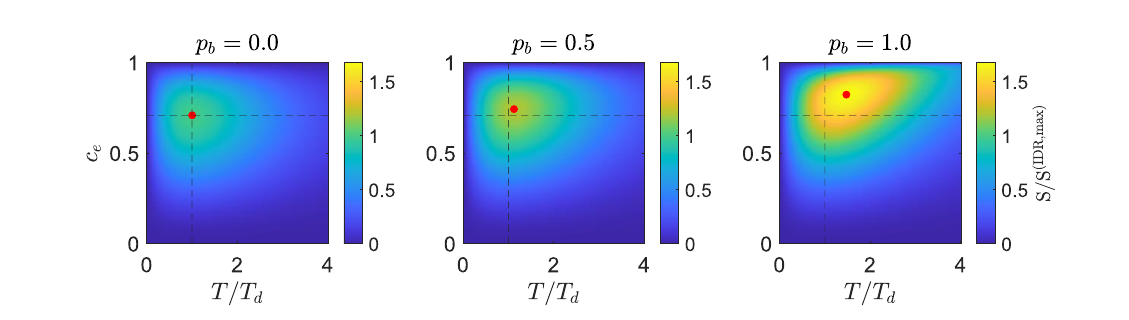}
    \caption{Sensitivity of a single atom Ramsey with decay detection for various experimental parameters. Here, we show the difference functional dependence on interrogation parameters $\frac{T}{T_{d}}$ and $c_{e}=\sqrt{1-c_{g}^{2}}$ for decay probabilities $p_{b}=0.0.5,1$. Sensitivity is scaled by the maximal IDR sensitivity $\text{S}^{\left(\text{IDR, max}\right)}=\frac{\tau T_{d}}{e}$. Red circle marks the optimal sensitivity, and dashed black lines indicate the optimal parameters for $p_{b}=0$, which is equivalent to the two-level model.}
    \label{fig: single atom sensitivity for various parameters}
\end{figure}

In order to obtain the sensitivity the measurement variance originating from projection noise must be calculated, using the successful interrogations number $N\left(c_{e},T\right)=\frac{\tau}{T}\left(1-p_{b}c_{e}^{2}\left(1-e^{-\Gamma T}\right)\right)$:
\begin{equation}
    \begin{split}
        \sigma\left(c_{g},\delta,T,\phi\right)&=\frac{1}{\frac{\tau}{T}\left(1-p_{b}c_{e}^{2}\left(1-e^{-\Gamma T}\right)\right)}\left(\frac{1}{2}+\frac{c_{e}c_{g}e^{-\frac{\Gamma}{2}T}\cos\left(\delta T-\phi\right)}{1-p_{b}c_{e}^{2}\left(1-e^{-\Gamma T}\right)}\right)\left(\frac{1}{2}-\frac{c_{e}c_{g}e^{-\frac{\Gamma}{2}T}\cos\left(\delta T-\phi\right)}{1-p_{b}c_{e}^{2}\left(1-e^{-\Gamma T}\right)}\right)\\
        &=\frac{\frac{1}{4}-\left(\frac{c_{e}c_{g}e^{-\frac{\Gamma}{2}T}\cos\left(\delta T-\phi\right)}{1-p_{b}c_{e}^{2}\left(1-e^{-\Gamma T}\right)}\right)^{2}}{\frac{\tau}{T}\left(1-p_{b}c_{e}^{2}\left(1-e^{-\Gamma T}\right)\right)}.
    \end{split}
\end{equation}
Finally, the sensitivity is given by:
\begin{equation}
    \begin{split}
        &\text{S}=\left.\frac{\left(\frac{\partial}{\partial\delta}P_{e| \text{not }g_{b}}\left(c_{g},\delta,T,\phi\right)\right)^{2}}{\sigma^{2}\left(c_{g},\delta,\tau,T,\phi\right)}\right|_{\delta T-\phi=\frac{\pi}{2}+n\pi}=4\tau T\frac{c_{e}^{2}c_{g}^{2}e^{-\Gamma T}}{1-p_{b}c_{e}^{2}\left(1-e^{-\Gamma T}\right)}.
    \end{split}
\end{equation}
We define the decay time $T_{d}$ and write the sensitivity using the unitless parameter $\frac{T}{T_{d}}$:
\begin{equation} \label{eq: single atom sensitivity}
    \begin{split}
        &\text{S}\left(c_{e},\frac{T}{T_{d}}\right)=4\tau T_{d}\frac{\frac{T}{T_{d}}c_{e}^{2}c_{g}^{2}e^{-\frac{T}{T_{d}}}}{1-p_{b}c_{e}^{2}\left(1-e^{-\frac{T}{T_{d}}}\right)}.
    \end{split}
\end{equation}
The sensitivity for various values of $\frac{T}{T_{d}}$ and $c_{g}=\sqrt{1-c_e^{2}}$ for few different decay probabilities $p_{b}$ is shown in Fig. \ref{fig: single atom sensitivity for various parameters}.

\section{Mid-interrogation single decay measurement}

In this section we calculate the sensitivity of a Ramsey measurement, where, in addition to the decay detection at the end of the interrogation, we incorporate a mid-interrogation decay measurement $\text{M}_{1}$ at time $rT$, where $0\le r\le1$ and $T$ is the full interrogation time. Unlike the continuous case, here the optimization of the sensitivity includes the additional parameter $r$.
 
 \subsection{Single atom with single mid-interrogation decay measurement}
 
\begin{figure}[h]
    \centering
    \includegraphics[width=1\linewidth]{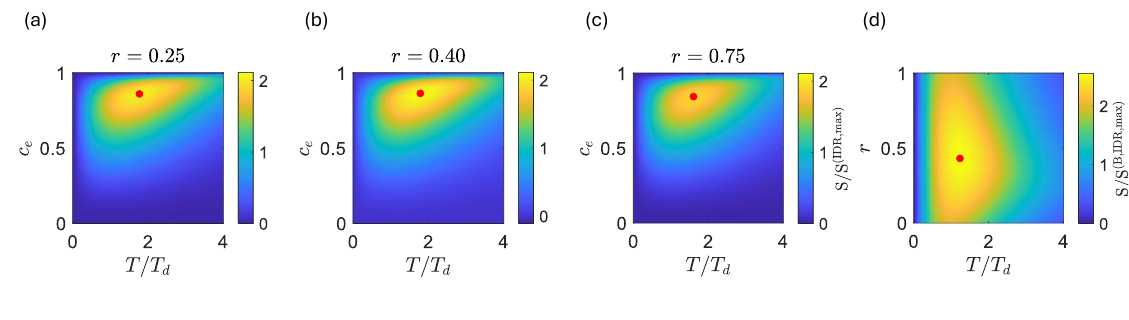}
    \caption{Sensitivity with single mid-interrogation decay for various experimental parameters. (a--c) Single atom sensitivity as a function of the interrogation parameters $\frac{T}{T_{d}}$ and $c_{e}=\sqrt{1-c_{g}^{2}}$ with decay detection at $rT$ for the optimal $r\approx0.4$ and examples at $r=0.25$ and $r=0.75$. The sensitivity is scaled by the maximal IDR sensitivity $\text{S}^{\left(\text{IDR, max}\right)}=\frac{\tau T_{d}}{e}$. (d) Bell state sensitivity as a function of the interrogation parameters $\frac{T}{T_{d}}$ and $r$, setting the mid-interrogation detection time at $rT$. Sensitivity is scaled by the maximal Bell state IDR sensitivity $\text{S}^{\left(\text{B, IDR, max}\right)}=\frac{\tau T_{d}}{2e}$. Red circle marks the optimal sensitivity.}
    \label{DDR mid single figure}
\end{figure}

We begin with the probability of a single atom to decay to $g_b$ between time $t'$ and $t'+dt'$:
\begin{equation}\label{eq: supple dP_{g_b}}
    dP_{g_{b}}\left(c_{e},t'\right)=p_bc^{2}_{e}e^{-\Gamma t'}\Gamma dt'.
\end{equation}
Integrating \ref{eq: supple dP_{g_b}}, the probability of the atom to decay between the interrogation start time $t=0$ and the time $t=rT$ is
\begin{equation}
    P_{g_{b}}\left(c_{e},rT\right)=\int_{0}^{rT}p_bc_{e}^{2}e^{-\Gamma t'}\Gamma dt'=p_bc_{e}^{2}\left(1-e^{-\Gamma rT}\right).
\end{equation}
The mean interrogation time for the single mid-interrogation decay detection case is calculated using $P_{g_{b}}\left(c_{e},rT\right)$ as
\begin{equation}\label{eq: T_mean_single}
\begin{split}
    &T_{\text{mean, single}}=P_{g_{b}}\left(c_{e},rT\right)rT+\left(1-P_{g_{b}}\left(c_{e},rT\right)\right)T=T\left[1-p_b c_{e}^{2}\left(1-e^{-\Gamma rT}\right)\left(1-r\right)\right].
\end{split}
\end{equation}
We note that $T_{\text{mean, single}}=T$ for both $r=0$ and $r=1$, as for the former there is no chance of decay by the time $rT=0$ and for the latter the mid-interrogation decay measurement time is $T$, meaning all interrogations are with duration $T$.
The sensitivity in this case is calculated using the successful measurements number
\begin{equation}
    \begin{split}
        N^{\left(\text{mid, single}\right)}\left(c_{e},T,r\right) &=\frac{\tau}{T_{\text{mean, single}}}\left(1-p_bc_e^2(1-e^{-\Gamma T})\right) = \frac{\tau\left(1-p_bc_e^2(1-e^{-\Gamma T})\right)}{T\left[1-p_b c_{e}^{2}\left(1-e^{-\Gamma rT}\right)\left(1-r\right)\right]} \\
        S^{\text{(mid,single)}}\left(c_e,\frac{T}{T_d},r\right) &= \frac{T}{T_{\text{mean,single}}} S\left(c_e,\frac{T}{T_d},r\right).
    \end{split}
\end{equation}
In the DDR case, this reads
\begin{equation}
    \text{S}^{\left(\text{DDR, mid, single}\right)}\left(c_{e},\frac{T}{T_{d}},r\right)=4\tau T_{d}\frac{\frac{T}{T_{d}}c_{e}^{2}c_{g}^{2}e^{-\frac{T}{T_{d}}}}{\left(1-c_{e}^{2}\left(1-e^{-r\frac{T}{T_{d}}}\right)\left(1-r\right)\right)\left(c_{g}^{2}+c_{e}^{2}e^{-\frac{T}{T_{d}}}\right)}.
\end{equation}
Optimizing $\text{S}^{\left(\text{DDR, mid, single}\right)}\left(c_{e},\frac{T}{T_{d}},r\right)$ we get the sensitivity, 
\begin{equation}
    \text{S}^{\left(\text{DDR, mid, single, max}\right)}\approx2.1\frac{\tau T_{d}}{e},
\end{equation}
with the parameters $c_e^{\text{(DDR, mid, single, max)}}\approx0.86$, $T^{\text{(DDR, mid, single, max)}} \approx 1.79~T_d$, and $r^{\text{(DDR, mid, single, max)}} \approx 0.40$. The sensitivity $\text{S}^{\left(\text{DDR, mid, single}\right)}\left(c_{e},\frac{T}{T_{d}},r\right)$ is shown in Figs. \ref{DDR mid single figure}a, \ref{DDR mid single figure}b and \ref{DDR mid single figure}c as a function of $c_{e}$ and $\frac{T}{T_{d}}$ for the optimal $r^{\left(\text{DDR, mid, single, max}\right)}$ and additional examples at $r=0.25,0.75$. 

\subsection{Bell state with single mid-interrogation decay measurement}

We begin with the probability of one of the atoms in a Bell state to decay between time $t'$ and $t'+dt'$:
\begin{equation}\label{eq: d_P_{g_{b}}^{(B)}}
    dP^{\left(\text{B}\right)}_{g_{b}}\left(t'\right)=e^{-\Gamma t'}\Gamma dt'.   
\end{equation}
Integrating Eq. \ref{eq: d_P_{g_{b}}^{(B)}} between the interrogation start time $t=0$ and $rT$ we obtain the probability of one of the atoms to decay to $\left|g_{b}\right\rangle$ within that time interval,
\begin{equation}
    P_{g_{b}}^{\left(\text{B}\right)}\left(rT\right)=\int_{0}^{rT}e^{-\Gamma t'}\Gamma dt'=1-e^{-\Gamma rT}.
\end{equation}
We calculate the mean interrogation time using $P_{g_{b}}^{\left(\text{B}\right)}\left(rT\right)$ as
\begin{equation}
    T_{\text{mean, single}}^{\left(\text{B}\right)}=\left(1-e^{-\Gamma rT}\right)rT+\left(1-\left(1-e^{-\Gamma rT}\right)\right)T=T\left[r+e^{-\Gamma rT}\left(1-r\right)\right].
\end{equation}
As in the single atom case, $T_{\text{mean, single}}^{\left(\text{B}\right)}=T$ for $r=0$ and $r=1$. The probability for a successful interrogation of time $T$ is $e^{-\Gamma T}$, and therefore the sensitivity in this case is calculated using the successful measurements number
\begin{equation}
    \begin{split}
        N^{\left(\text{B, DDR, mid, single}\right)}\left(T,r\right)=\frac{\tau}{T_{\text{mean, single}}^{\left(\text{B}\right)}}e^{-\Gamma T}=\frac{\tau e^{-\Gamma T}}{T\left[r+e^{-\Gamma rT}\left(1-r\right)\right]},
    \end{split}
\end{equation}
and reads:
\begin{equation}
    \text{S}^{\left(\text{DDR, mid, single}\right)}\left(\frac{T}{T_{d}},r\right)=\tau T_{d}\frac{\frac{T}{T_{d}}e^{-\frac{T}{T_{d}}}}{r+e^{-r\frac{T}{T_{d}}}\left(1-r\right)}.
\end{equation}
Optimizing $r$ and $\frac{T}{T_{d}}$ simultaneously, we obtain the maximal sensitivity 
\begin{equation}
    \text{S}^{\left(\text{B, DDR, mid, single, max}\right)}\left(\frac{T}{T_{d}},r\right)\approx2.55\frac{\tau T_{d}}{2e}.
\end{equation}
Fig. \ref{DDR mid single figure}d depicts the sensitivity $\text{S}^{\left(\text{B, DDR, mid, single}\right)}\left(\frac{T}{T_{d}},r\right)$ as a function of its parameters.

\section{Information-theoretic bounds}
\subsection{Quantum Fisher information}
    \par The gain in sensitivities presented above can be equivalently calculated by also considering the quantum Fisher information (QFI) $\mathcal{F}(\rho)$. The QFI characterises the maximum information about a parameter $\theta$ that can be extracted from a quantum state $\rho(\theta)$ over any possible measurement strategy. Formally, the QFI bounds the achievable precision of any locally unbiased estimator $\hat{\theta}$ by the Quantum Cram\'er-Rao Bound (QCRB):
    \begin{equation}
        \cov{\hat{\theta}} \geq \frac{1}{N_{tot}}\mathcal{F}^{-1},
    \end{equation}
    where $N_{tot}$ is the number of copies of $\rho$ available. 
    
    \par The QFI can be evaluated from the spectral decomposition $\rho = \sum_i \lambda_i \dm{\lambda_i}$.
    \begin{equation}
    \begin{split}
    \label{eqn:QFI}
        \mathcal{F} = &\sum_{\lambda_i\neq0} \left(\frac{(\partial_a \lambda_i)^2}{\lambda_i} + 4 \lambda_i \braket{\partial_a \lambda_i}{\partial_a \lambda_j} \right) \\
        - &\sum_{\lambda_i, \lambda_j \neq0} \frac{ 8 \lambda_i\lambda_j}{\lambda_i + \lambda_j} \abssq{\braket{\partial_a \lambda_i}{\lambda_j}}.
    \end{split}
    \end{equation}
    The density matrix (\ref{eq: supple  solution to Lindblad equation}) can thus be diagonalised to get the QFI for a single shot $\mathcal{F}_{shot
    }$. In practice, we want to optimize the QFI for wall-clock time -- going to longer probe durations may give more information per shot, but at the expense of doing fewer shots in the same time. We thus evaluate the average QFI in terms of the wall-clock time $\tau$: 
    \begin{align}
        \mathcal{F}(c_e, T) &= \frac{\tau}{T} \mathcal{F}_{shot} \\
        &= 4\tau T_d  \frac{\frac{T}{T_d}c_e^2 c_g^2 e^{-\frac{T}{T_d}}}{1 -p_b c_e^2(1-e^{-\frac{T}{T_d}})}. \label{eqn:css-qfi}
    \end{align}
    This is exactly the sensitivity derived in Eq. \ref{eq: single atom master equation}, proving that the measurement protocol described is information-theoretic optimal. 

    In the presence of individual dephasing on each atom with rate $\gamma$, the QFI is modified to $\mathcal{F}_{dephased} = e^{-\gamma t} \mathcal{F}$, the discussion of which is outside the scope of this work.

    \subsection{Optimal parameters}
    We can optimize the QFI in the initial excitation for a given probe duration:
    \begin{align}
        c_{e}^{\text{max}
        }(T)^2 &= \frac{1}{1+\sqrt{p_a + p_b e^{-
        \Gamma T}}} \label{eqn:ghz-optimal-a}\\
        \mathcal{F}(c_{e}^{\text{max}}, T) &= 4 \tau T e^{-\Gamma T} c_{e}^{\text{max}}(T)^4 \label{eqn:qfi-optimal-ce}.
    \end{align}
    For general branching ratios $p_a$, Eq. \eqref{eqn:qfi-optimal-ce} must be maximized numerically to find the optimal probe duration. In the DDR limit, the optimal probe duration has the closed-form expression
    \begin{equation}
        \left(\frac{T}{T_d}\right)^{(\text{DDR,max})}= 1 + 2W_0\left(\frac{1}{\sqrt{2}e}\right) \approx 1.48,
    \end{equation}
    where $W_0(z)$ is the principal branch of the Lambert W function $W_k(z)$, which is the solution to the equation $W_k(z)e^{W_k(z)}=z$. The corresponding initial excitation and QFI are
    \begin{align}
        \left(c_e^{(\text{DDR,max})}\right)^2&=\frac{1}{1+2 W_0\left(\frac{1}{\sqrt{2}e}\right)} \approx 0.68, \\
        \mathcal{F}^{(\text{DDR,max)}}&= \tau T_d \frac{16 W_0\left(\frac{1}{\sqrt{2}e}\right)^2}{1 + 2 W_0\left(\frac{1}{\sqrt{2}e}\right)}  \approx 1.68 \frac{~\tau T_d}{e}.
    \end{align}
    It is interesting to note that the populations of the excited state $\ket{e}$ and the clock ground state $\ket{g_a}$ are equal at \emph{half} the optimal probe duration, and thus their relative populations are inverted at the end of the interrogation sequence. 

    These results extend trivially to the $n-$atom coherent spin state (CSS). The QFI scales linearly in the number of atoms $\mathcal{F}_{\text{CSS}}= n \mathcal{F}$, and the optimal parameters remain unchanged.

\section{Extension to GHZ states}
    \par Now consider the imbalanced $n$-atom Greenberger–Horne–Zeilinger (GHZ) state $c_e\ket{e}^{\otimes n} + c_g \ket{g_a}^{\otimes n}$. The master equation Eq. \eqref{eq: single atom master equation} is slightly modified to include the dynamics of all $n$ atoms:
    \begin{equation} \label{eq: n atom master equation}
        \frac{\partial}{\partial t}\rho=-i \sum_{k=1}^n\left(\left[S_{z}^{(k)},\rho\right]+\frac{\Gamma}{2}\mathcal{L}_{\Lambda_{a}^{(k)}}\left[\rho\right]+\frac{\Gamma}{2}\mathcal{L}_{\Lambda_{b}^{(k)}}\left[\rho\right] \right),
    \end{equation}
    where $A^{(k)}$ is the single-atom operator $A$ acting on the $k^{\text{th}}$ atom. After free-evolution for duration $t$ governed by (\ref{eq: n atom master equation}), the GHZ state evolves to
    \begin{align}
        \begin{split}
        \label{eqn:ghz-evolved}
        \rho_{\text{GHZ}}(\delta, t)= \: & c_g^2 \dm{g_a}^{\otimes n} + c_e c_g e^{-\frac{n\Gamma}{2} t} \left[e^{-i n\delta t} \ketbra{e}{g_a}^{\otimes n} + \text{h.c.} \right] \\
        & + c_e^2\left[e^{-\Gamma t} \dm{e}^{\otimes n} + \left(1-e^{-\Gamma t}\right) \Bigl(p_a \dm{g_a} + p_b\dm{g_b}\Bigr) \right]^{\otimes n}.
        \end{split}
    \end{align}

    \par
\begin{figure}
        \centering
        \subfloat[]{
        \label{sfig:GHZ optimal qfi}
        \includegraphics[width=0.45\textwidth]{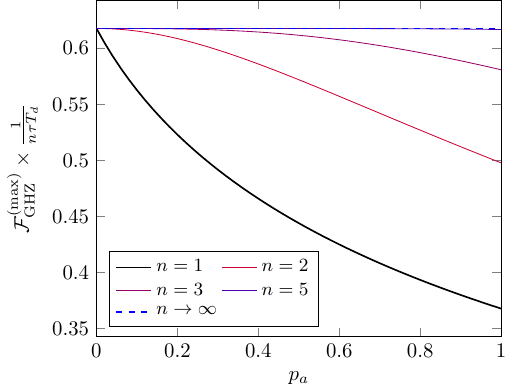}
        } \hfill
        \subfloat[]{
        \label{sfig:GHZ optimal probe duration}
        \includegraphics[width=0.45\textwidth]{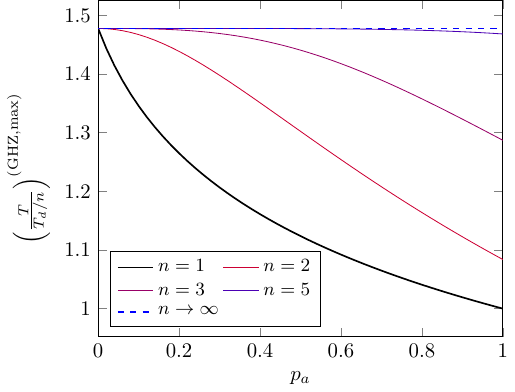}
        }
        \caption{\protect\subref{sfig:GHZ optimal qfi} Optimal quantum Fisher information (QFI) per atom, and \protect\subref{sfig:GHZ optimal probe duration} the corresponding probe duration (normalized to the effective lifetime $T_d/n$) as a function of the excited state branching ratio to the clock ground state $g_a$, plotted for GHZ states of various sizes $n$. As $n$ increases, both the QFI and the optimal duration become independent of $p_a$ and asymptote to the single-atom DDR limit.}
        \label{fig:GHZ-css comparison}
    \end{figure}

 Outside the GHZ subspace spanned by $\{ \ket{e}^{\otimes n}, \ket{g_a}^{\otimes n} \}$, the evolved state is already diagonal and carries no information about the laser frequency. Within the GHZ subspace, the density matrix $\rho_{\text{GHZ}}(\delta,t)$ can then be easily diagonalised to evaluate the QFI
    \begin{equation}
        \mathcal{F}_{\text{GHZ}}(c_e, T) = 4 \tau T_d n^2 \frac{ \frac{T}{T_d} c_e^2 c_g^2 e^{-n \frac{T}{T_d}}}{1 - c_e^2 \left(1 - e^{-n \frac{T}{T_d}} - p_a^n \left(1-e^{-\frac{T}{T_d}}\right)^n\right)},
        \label{eqn:ghz-qfi}
    \end{equation}
    such that the optimal initial excitation is
    \begin{equation}
        \left(c_e^{(\text{GHZ,max)}}(T)\right)^2 = \frac{1}{1+\sqrt{e^{-n \Gamma T}+p_a(1-e^{-\Gamma T})^n}}.
    \end{equation}
    Fig. \ref{fig:GHZ-css comparison} plots the optimal QFI and probe duration as a function of the decay branching ratio. In the IDR case ($p_a=1$), Eq. \eqref{eqn:ghz-qfi} reduces to the QFI derived in \cite{kielinski2024ghz}.

    \par In the DDR case, we see that $\mathcal{F}_{\text{CSS}}^{\text{(DDR)}}(c_e, T) = \mathcal{F}_{\text{GHZ}}^{\text{(DDR)}}(c_e, T/n)$. This directly implies that the CSS and GHZ states have exactly the same information-theoretic limits, with $T^{(\text{DDR,max})}_{GHZ} = T^{(\text{DDR,max})}_{CSS}/n$ -- i.e., entanglement provides no enhancement. This can be attributed to two factors: 
    \begin{enumerate*}[label=(\roman*)]
        \item the GHZ state is more sensitive to spontaneous decay, so the effective decay rate increases $\Gamma_{\text{GHZ}}=n\Gamma$, and
        \item all decay events can be detected with certainty in both the CSS and GHZ states, so unsuccessful interrogations are discarded with the same fidelity in both cases.
    \end{enumerate*}.
    
    \par As the probability of decaying to clock ground state $p_a$ increases, the latter no longer holds and the entangled states do provide an advantage (Fig. \ref{sfig:GHZ optimal qfi}). With a single atom, one can only discard decays that occur outside the clock subspace. The GHZ state, however, can additionally distinguish interrogations where some (but not all) atoms decay to $g_a$. As a result, the probability of incorrectly accepting a decayed state is $c_e^2 p_a (1-e^{-\Gamma T})$ with a single atom, but only $c_e^2 p_a^n (1-e^{-\Gamma_{\text{GHZ}} T/n})^n$ for a GHZ state. For $n \gg 1$, this is vanishingly small for all $p_a$, and the GHZ state QFI becomes independent of the branching ratio. The DDR limit thus applies to any Ramsey-style interrogation in which a spontaneous decay can be deterministically detected, explaining the equivalence to the results in \cite{kielinski2024ghz}. 
     
    \subsection{Mid-interrogation measurements}
    Similarly to the single-atom Ramsey case, we can calculate the mean interrogation time for mid-interrogation $M_1$ measurement in both the continuous limit and for single mid-interrogation measurement at time $rT$:
    \begin{align}
        T_{\text{mean, cont.}}^{(\text{GHZ})} &= 1-p_b c_e^2 + \frac{c_e^2}{\Gamma T} \left(\frac{1-e^{-n \Gamma T}}{n} - p_a^n \sum_{m=1}^{n} \frac{\left(1-e^{- \Gamma T}\right)^m}{m}  \right), \\
        T_{\text{mean, single}}^{(\text{GHZ})} &= c_g^2 + (1-r) c_e^2 \left( e^{-n\Gamma rT} +p_a^n (1-e^{-\Gamma rT})^n\right),
    \end{align}
    with the QFI scaled up as $\mathcal{F}_{\text{GHZ}}^{\text{(mid)}}(c_e,T) = \frac{T}{T_{\text{mean}}^{\text{(GHZ)}}} \mathcal{F}_{\text{GHZ}}(c_e,T)$.

    \subsection{Practical considerations}
    \par The optimal measurement strategy to saturate the QFI for the GHZ state follows the same idea as the single atom case. After free evolution, the measurement $M_1$ is first performed on all atoms to discard any decays to the third state $g_b$. The measurement $M_2$, however, is modified to be the ``heralded-uGHZ protocol" described in~\cite{kielinski2024ghz}, which uses a one-axis twisting (OAT) operation to identify and discard decays to $g_a$. If we limit ourselves to using only single-qubit interactions for measurement, the advantage for $p_a>0$ is lost and the GHZ state becomes only as sensitive as the CSS. In practice, it may be simpler to use a CSS on a transition with $b=0$ than to implement a finely controlled OAT for readout.
    
\bibliography{stability_limit_paper_references.bib}